\documentclass[twocolumn]{aastex63}

\newcommand{\alphaox}{{$\alpha_{\rm ox}$\,}}

\newcommand{\boleddfrac}{{f${_{\rm Edd, bol}}$\,}}

\submitjournal{ApJ}

\shorttitle{Accretion state transitions in a TDE}
\shortauthors{Wevers et al.}

\graphicspath{{./}{figures/}}
\begin{document}

\title{Rapid accretion state transitions following the tidal disruption event AT2018fyk}

\correspondingauthor{Thomas Wevers}
\email{twevers@eso.org}
\author[0000-0002-4043-9400]{T. Wevers}
\affiliation{European Southern Observatory, Alonso de Córdova 3107, Vitacura, Santiago, CL}
\affiliation{Institute of Astronomy, University of Cambridge, Madingley Road, CB3 0HA, Cambridge, UK}

\author{D.R. Pasham}
\affiliation{MIT Kavli Institute for Astrophysics and Space Research, Cambridge, MA 02139, USA}

\author{S. van Velzen}
\affiliation{Leiden Observatory, Leiden University, P.O. Box 9513, 2300 RA, Leiden, NL}
\affiliation{Center for Cosmology and Particle Physics, New York University, NY 10003, USA}
\affiliation{Department of Astronomy, University of Maryland, College Park, MD 20742, USA}

\author{J.C.A. Miller-Jones}
\affiliation{ICRAR -- Curtin University, GPO Box U1987, Perth, WA 6845, AU}

\author{P. Uttley}
\affiliation{Anton Pannekoek Institute, University of Amsterdam, Science Park 904, 1098 XH, Amsterdam, NL}

\author{K.C. Gendreau}
\affiliation{Astrophysics Science Division, NASA Goddard Space Flight Center, Greenbelt, MD 20771, USA}
\author{R. Remillard}
\affiliation{MIT Kavli Institute for Astrophysics and Space Research, Cambridge, MA 02139, USA}

\author{Z. Arzoumanian}
\affiliation{Astrophysics Science Division, NASA Goddard Space Flight Center, Greenbelt, MD 20771, USA}

\author{M. L\"{o}wenstein}
\affiliation{Astrophysics Science Division, NASA Goddard Space Flight Center, Greenbelt, MD 20771, USA}

\author{A. Chiti}
\affiliation{MIT Kavli Institute for Astrophysics and Space Research, Cambridge, MA 02139, USA}

\begin{abstract}
Following a tidal disruption event (TDE), the accretion rate can evolve from quiescent to near-Eddington levels and back over months--years timescales. This provides a unique opportunity to study the formation and evolution of the accretion flow around supermassive black holes (SMBHs). We present two years of multi-wavelength monitoring observations of the TDE AT2018fyk at X-ray, UV, optical and radio wavelengths. We identify three distinct accretion states and two state transitions between them. These appear remarkably similar to the behaviour of stellar-mass black holes in outburst. The X-ray spectral properties show a transition from a soft (thermal-dominated) to a hard (power-law dominated) spectral state around L$_{\rm bol} \sim $few$ \times 10^{-2}$ L$_{\rm Edd}$, and the strengthening of the corona over time $\sim$100--200 days after the UV/optical peak. Contemporaneously, the spectral energy distribution (in particular, the UV-to-X-ray spectral slope \alphaox) shows a pronounced softening as the outburst progresses. The X-ray timing properties also show a marked change, initially dominated by variability at long ($>$day) timescales while a high frequency ($\sim$10$^{-3}$ Hz) component emerges after the transition into the hard state. At late times ($\sim$500 days after peak), a second accretion state transition occurs, from the hard into the quiescent state, as identified by the sudden collapse of the bolometric (X-ray+UV) emission to levels below 10$^{-3.4}$ L$_{\rm Edd}$. Our findings illustrate that TDEs can be used to study the scale (in)variance of accretion processes in individual SMBHs. Consequently, they provide a new avenue to study accretion states over seven orders of magnitude in black hole mass, removing limitations inherent to commonly used ensemble studies.
\end{abstract}

\keywords{tidal disruption events --- accretion disks --- black holes --- active galactic nuclei}

\section{Introduction} \label{sec:intro}
Stellar-mass black holes (with masses between 5--15 solar masses [$M_{\odot}$]) can produce outbursts of radiation triggered by a sudden influx of material. The emission from these outbursts encodes critical information about the physics of accretion onto a compact object, nature's most efficient way to convert mass to energy. Broadly speaking, three accretion states are observed during these outbursts: the quiescent, the soft, and the hard state \citep{Homan05,Remillard06}. These states are defined by the relative importance of the two main physical components of the inner accretion flow: the accretion disk, and a hot, tenuous plasma known as the corona.

In quiescence and at the start of an outburst, stellar-mass black holes typically produce relatively hard emission (dominated by higher energy photons). As it nears peak brightness, the emission becomes soft, i.e., dominated by relatively lower energy photons from the accretion disk, with a weak coronal contribution and low X-ray flux variability. This state is referred to as the soft state. As the outburst intensity declines, the emission becomes harder again, dominated by higher energy X-ray photons from the corona. This hard state is also characterised by large X-ray variability. As the brightness decreases further, these systems return to the quiescent state. This last is poorly understood because the system is intrinsically very faint, but quiescent emission is typically softer than in the hard state. Evolution across states, which represents a framework to understand the process of accretion onto black holes, is thought to correlate with the overall mass accretion rate, although other factors (e.g. black hole spin, magnetic flux) likely also play an important role.

A long-standing question in compact object accretion physics is whether supermassive black holes (SMBHs; with masses $\gtrsim$ a few$\times$10$^{5}$ $M_{\odot}$) undergo the same accretion cycle, i.e., with similar states and mechanisms that trigger transitions between the states. This is interesting because state transitions signal changes to the fundamental physics governing the accretion processes \citep{Abramowicz13}. These include the dominant cooling mechanism, geometry, interplay between the emitting regions, jet formation and accretion efficiency. A key question is if and how these mechanisms scale with the black hole mass.

For actively accreting SMBHs or active galactic nuclei (AGN), large changes in mass accretion rate similar to stellar-mass black hole outbursts are thought to occur on timescales of hundreds to thousands of years. Observations of multiple state transitions in individual AGN are therefore very rare \citep{Mcelroy16, Parker19}, and the observed timescales ($\sim$ 10s of years) highlight the uncertainties in our understanding of the mechanism responsible for such state changes. Observational constraints for SMBHs in different accretion states are largely statistical in nature (\citealt{Merloni03, Falcke04, Mchardy06}, although some work on individual systems exists -- e.g. \citealt{Gezari17clagn, Noda18, Frederick19, Trakhtenbrot19}), which severely complicates the development of a detailed, holistic theoretical framework.

Tidal disruption events (TDEs) are episodes in which passing stars are ripped apart by tidal forces in the vicinity of SMBHs within distant galaxies \citep{Hills1975,Rees1988}. They have long been heralded as ideal systems to study accretion states and transitions in SMBHs, as they evolve from a dormant (quiescent) state to a high accretion rate phase, and back again, on $\sim$year timescales. However, many TDE accretion disks appear to be stable for at least 5--10 yr after disruption \citep{vanvelzen2019}. Moreover, only a small fraction of TDEs show persistently bright X-ray emission, without which a direct comparison to X-ray binary (XRB) spectral and timing properties is challenging. As a result, TDEs with multiple state transitions have not yet been found (see \citealt{Komossa04, Maksym14, Jonker20, Wevers20} for work in this regard).

The transient ASASSN--18ul/AT2018fyk was discovered by the All-Sky Automated Survey for Supernovae \citep{Shappee14} on 8 September 2018 in the nucleus of a galaxy at a redshift of 0.059 (luminosity distance of 264 Mpc). Based on the blue optical spectrum with broad H and He emission lines, hot (T$\sim$35000 K) UV/optical blackbody emission that does not cool significantly over time, and the lack of AGN-like emission lines, it was classified as a TDE by a black hole with a (independently derived) mass of log$_{10}$(M$_{\rm BH}$) = 7.7 $\pm$ 0.4 M$_{\odot}$\citep{Wevers20, Wevers19b}.\\

In this manuscript we present an in-depth analysis of both archival and new radio, optical, UV and X-ray observations of AT2018fyk taken up to 2 years after the initial discovery. These data provide three key diagnostics to characterise the accretion flow properties following the TDE, which together enable detailed comparison with stellar-mass black holes:
\begin{itemize}
    \item The evolution of the UV--X-ray spectral slope \alphaox with bolometric Eddington ratio.
    \item The ratio of the power-law spectral component flux to the total emission in X-rays, and the ratio of power-law emission to the total bolometric luminosity (including the accretion disk, soft excess and corona).
    \item Photometric variability properties of the X-ray emission derived from the {\it Swift} and {\it XMM-Newton} light curves.
\end{itemize} 
Using these diagnostic tools, we investigate the properties of AT2018fyk, and find remarkable similarities to the properties of accreting stellar-mass black holes.\\

Section \ref{sec:reduction} details the observations and data reduction. We present the results of host galaxy modelling and SED, X-ray energy spectral and timing analysis in Section \ref{sec:analysis}. The main results are discussed in the framework of accretion state transitions in Section \ref{sec:discussion}, including similarities and differences between stellar-mass and SMBH systems, and a comparison between AT2018fyk and AGNs. We summarise our findings in Section \ref{sec:summary}.

\section{Observations and data reduction}
\label{sec:reduction}
\subsection{XMM-Newton stare observations}
AT2018fyk was observed by the {\it XMM-Newton} \citep{Jansen01} European Photon Imaging Camera (EPIC) on three occasions. The first observation (obsID: 0831790201; hereafter referred to as XMM1), for a total of 32 kiloseconds (ks), was observed shortly after the initial disk transition began (between states A and B) on 2018 December 9 (MJD 58461.75), $\approx$ 92 days after discovery. 
The second observation (obsID: 0853980201; XMM2) was performed $\approx$ 413 days after discovery (in state D), for a total of 54 ks on 2019 October 26 (MJD 58782.21). 
The third observation, totalling 17 ks (obsID: 0854591401; XMM3), was taken on 2020 May 12 (MJD 58981.27), roughly 612 days after optical discovery and after the second state transition (in state E). 
In all the observations EPIC was operating in the imaging mode and the source did not suffer from pile-up\footnote{\url{https://heasarc.gsfc.nasa.gov/docs/xmm/sas/USG/epicpileup.html}}.

We visually inspect the EPIC (pn+MOS) images from all three observations. A point source coincident with the optical position ($\alpha$=22:50:16.090, $\delta$=--44:51:53.50, \citealt{Wevers19b}) is present in both XMM1 and XMM2 images. However, no source is visible in the X-ray image of XMM3.

We start our data reduction with the raw Observation Data Files (ODFs) and process them using the {\it XMM-Newton} Standard Analysis System (SAS) version 17.0.0 tools {\tt epproc} and {\tt emproc}. Using the most recent version of the calibration database (CALDB) files this procedure results in ``cleaned'' event files, which are used for deriving scientific products. We extract good time intervals (GTIs) for each of the three observations by screening for periods of background flaring and considering only times when all detectors (pn+MOS1+MOS2) are active. 

Source events are extracted from a circular aperture of radius 33 arcsec, which corresponds to roughly 90\% of light as estimated from the encircled energy function of the EPIC instruments. For each observation, background events are extracted from two circular regions of radius 50 arcsec close to the source and on the same CCD. The energy spectra and light curves are corrected for the different source and background areas. Standard data filters of {\it (PATTERN $<=$ 4) } and {\it (PATTERN $<=$ 12)} are applied for the pn and MOS data, respectively; we only use events in the 0.3--10 keV band for further analysis. Corresponding response files were generated using {\tt XMM SAS} tools {\tt rmfgen} and {\tt arfgen}.

For the XMM3 observation, we estimate the X-ray flux upper limit (3$\sigma$) using the EPIC/pn data by following the SAS documentation\footnote{\url{https://www.cosmos.esa.int/web/xmm-newton/sas-thread-src-find}}. Background flaring during XMM3 reduced the effective exposure from 17 ks to about 10 ks. Running the {\it XMMSAS} tool {\tt edetect\_chain} on a 0.3--10 keV pn image does not yield a point source at the location of AT2018fyk. We estimate the 0.3--10 keV count rate upper limit at the source location to be 0.0065 counts s$^{-1}$ using the sensitivity map generated by the {\tt edetect\_chain} task. We then use the {\tt fakeit} tool in the X-ray spectral fitting program {\it XSPEC} \citep{Arnaud96} and the {\it XMM-Newton} response files generated using the {\tt arfgen} and {\tt rmfgen} commands to estimate the flux upper limit. Because no constraints on the X-ray spectral shape are available, we estimate the upper limit using the best fit spectral model parameters from XMM1 and XMM2 with mean 0.3--10 keV count rates of 0.87 and 1.13 counts s$^{-1}$, respectively (see Table \ref{tab:xrayspectra}). Assuming the flux scales linearly in an energy-independent manner, 0.3--10 (0.01--10) keV 3$\sigma$ upper limits are 8.4$\times$10$^{-15}$ (7.8$\times$10$^{-14}$) and 1.2$\times$10$^{-14}$ (2.2$\times$10$^{-14}$) erg cm$^{-2}$ s$^{-1}$ for XMM1 and XMM2, respectively. We adopt the more conservative estimates of the XMM2 spectral model in our analysis.

\subsection{Chandra stare observation}
Following the non detection in the latest {\it XMM-Newton} exposure (XMM3) we obtained a deep {\it Chandra}/ACIS-S Director’s Discretionary Time (DDT) observation. The 46.8 ks exposure started 660 days after the optical discovery (MJD 59029.22). For better sensitivity at low energies, we used the back-illuminated S3 chip in timed exposure mode with the telemetry set to very faint format. Similar to the {\it XMM-Newton}/EPIC data reduction, we start our ACIS analysis with the level-1 (secondary) data files which we reprocess using the {\it Chandra} data analysis software (CIAO version 4.10). We reduce the level-1 data with the {\tt chandra\_repro} script with default parameter values.

We analyse the level-2 event files by first extracting a broadband (0.5--7 keV) X-ray image of the field of view. We use the {\tt fluximage} task with a {\it binsize} of 1 (0.492 arcsec). Visual inspection of the resulting exposure-corrected image does not show a source at the optical position of AT2018fyk. Assuming Poisson statistics and using the CIAO task {\tt srcflux}, we estimate the 0.3--10 keV flux upper limit using the best fit {\tt bremss+pow} model parameters from XMM1 and XMM2. Both model parameters give a similar value for the unabsorbed 3$\sigma$ flux upper limit in the 0.3--10 keV energy range of 2$\times$10$^{-15}$ erg cm$^{-2}$ s$^{-1}$ .

\subsection{NICER monitoring observations}\label{sec:nicerdata}
The {\it NICER} \citep{Gendreau16} X-ray Timing Instrument (XTI) onboard the International Space Station (ISS) began monitoring AT2018fyk on MJD 58388.15; 242 observations were made with individual exposures lasting between a few tens and two thousand seconds. We consider all publicly available observations taken before 2020 January 13 (MJD 58861.34; up to obsID 2200370264). These data are processed using the standard {\it NICER} Data Analysis Software (NICERDAS) task {\tt nicerl2} with default filtering and the gain file {\it nixtiflightpi20170601v004.fits}. Consecutive observations are combined to ensure a minimum of 20 ks of cleaned exposure. This yields a total of 31 energy spectra over the entire monitoring period. 

All source spectra are extracted, and background spectra are estimated, as follows. Two empirical background spectrum libraries are constructed from observations of source-free areas of the sky. These include the cosmic X-ray (astrophysical) background associated with those fields as well as the {\it NICER}/XTI instrumental Non-X-ray-Background (NXB). The ISS night background library spectra are collected and categorised into discrete cells according to two background proxies. The first is the count rate of focused events in the 15--18 keV band, where the low effective area of the {\it NICER} optics assures that these are not X-rays from the source. The second is the rate of spatially extended (with respect to where X-rays are focused on the detector) events, as identified by their location in the plane of energy versus pulse invariant (PI) ratio of slow chain to fast chain detected events. These are the so-called ``trumpet’’ rejected  events\footnote{\url{https://heasarc.gsfc.nasa.gov/docs/nicer/mission$\_$guide/}}. Corresponding rates in the source data are extracted using the input cleaned event file GTIs, subdivided into
intervals no larger than 120 s, and a match to one of the background library cells identified. The total night background is calculated as a sum over the night library spectra, appropriately weighted by fractional exposure and scaled by the 15--18 keV count rate in each
interval relative to average value in the cell. A supplemental ISS day residual background library is constructed from the same source-free observation database to account for an additional background associated with an optical light leak. This is used to derive an additional background component identified and scaled according to the rate in the 0--0.25 keV energy band.

The total spectrum extracted from the input event list is based on these subdivided GTIs, with the additional excision of data in intervals with background proxies that lie outside of the background library bounds. Also, subdivided GTIs where the absolute value of the estimated net 13--15 keV count rate exceeds 0.1 counts per second, or the estimated background rate exceeds the total rate, are excluded. Data from noisy Focal Plane Modules (FPMs) 14 and 34 are filtered out, as well.

\subsection{Swift monitoring observations}
{\it Swift} started monitoring AT2018fyk on MJD 58383.69, just two weeks after its discovery. Between MJDs 58383.69 -- 59031.24 a total of 161 snapshots were made, which accumulated 198 ks of exposure. All these observations were performed as part of multiple Target of Opportunity (ToO) requests over the $\sim$650 day period. All the data used here are publicly available and can be downloaded from NASA's High Energy Astrophysics Science Archive (HEASARC).

To measure the UV and optical photometry, a circular aperture of 7" is used on images taken by the {\it Swift} UV and Optical Telescope (UVOT;  \citealt{Roming05}) images. We use the task \texttt{uvotsource} of HEASOFT v6.24. Sub-exposures of a given observation are combined before extracting the flux.  

For the X-ray Telescope (XRT) data, we start our analysis with the raw level-1 data. These are reduced using {\tt xrtpipeline} with the source optical coordinates. We only use data taken in photon counting mode with event grades between 0 and 12. Source events are extracted from a circular region with a radius of 47''. This value corresponds to roughly 90\% (at 1.5 keV) of the light from a point source (as estimated from the XRT fractional encircled energy function). Background events are extracted from an annular region with inner and outer radii of 70'' and 150'', respectively. These values were chosen to avoid point sources in the background annulus. Spectra are binned to a minimum of 20 counts per bin for spectral analysis.

X-ray light curves are then extracted using the {\tt xrtlccorr} task, which properly takes bad pixels and columns into account. For each exposure we estimate the source and background count rates separately. Finally, background corrected source rates are estimated by subtracting the area-scaled background rates from the source rates. These steps are repeated for both the soft (0.3--1.5 keV) and the hard (1.5--10 keV) X-ray bands to extract the soft and the hard X-ray light curves. 

To obtain a deeper constraint on the late time X-ray emission with {\it Swift}, we combine the individual exposures ranging from MJDs 58930 -- 58975. A 3$\sigma$ upper limit is derived on the combined image, taking into account the combined exposure map, using the {\it sosta} task in {\tt ximage}. The light curve data on a per obsID basis can be retrieved from the attached supplementary files ({\it XRT\_lightcurve\_30\_150.dat} and {\it XRT\_lightcurve\_150\_1000.dat}).

\subsection{Australia Telescope Compact Array observations}
\begin{table*}
    \centering
        \caption{ATCA observations of AT2018fyk. States are labelled as in Figure \ref{fig:alphaox}. Upper limits are reported for the stacked image made from the joint deconvolution of the 5.5 and 9.0\,GHz data. The corresponding luminosity upper limits are calculated assuming a central frequency of 7.25 GHz.}
    \begin{tabular}{ccccccc}
    \hline
State & MJD & Phase & On-source time & Config. & $3\sigma$ limit & L$_{\rm radio}$ \\
 & (days) & (MJD - 58369) & (min) & & ($\mu$Jy\,beam$^{-1}$) & (erg s$^{-1}$)\\
         \hline
 C &$58594.14$ & 225 & 63 & 750C & $<104$ & $<$6.3$\times$10$^{37}$\\ %
 C &$58646.99$ & 278 & 187 & 6A & $<27$ & $<$1.6$\times$10$^{37}$\\%
 E &$58952.20$ & 583 & 145 & 6A & $<46$ & $<$2.8$\times$10$^{37}$\\%
\hline
    \end{tabular}
    \label{tab:atca}
\end{table*}
Radio observations were obtained with the Australia Telescope Compact Array (ATCA). Observations were performed at early times (25, 52 and 89 days after discovery; see \citealt{Wevers19b} for details of these observations) as well as at late times (225, 278 and 583 days after discovery), as detailed in Table~\ref{tab:atca}.  For the new late time observations, we observed in the 4\,cm band to maximise sensitivity, with 2\,GHz of bandwidth in each of two frequency bands centred at 5.5 and 9.0\,GHz.  We used the standard calibrator source PKS 1934-638 to determine the band pass solutions and to set the amplitude scale.  We used the nearby calibrator source J2230-4416 to determine the complex gain solutions, which were interpolated onto the target. Data were reduced using the Common Astronomy Software Application (CASA; \citealt{McMullin2007}). AT2018fyk was not detected in either frequency band at any of the three late-time epochs, and we jointly deconvolved the data from both frequency bands to get the deepest possible upper limits on the source flux density, as reported in Table~\ref{tab:atca}.

\section{Analysis and results}
\label{sec:analysis}
\subsection{Host galaxy SED modelling}
\begin{figure}
\includegraphics[width=\columnwidth, keepaspectratio]{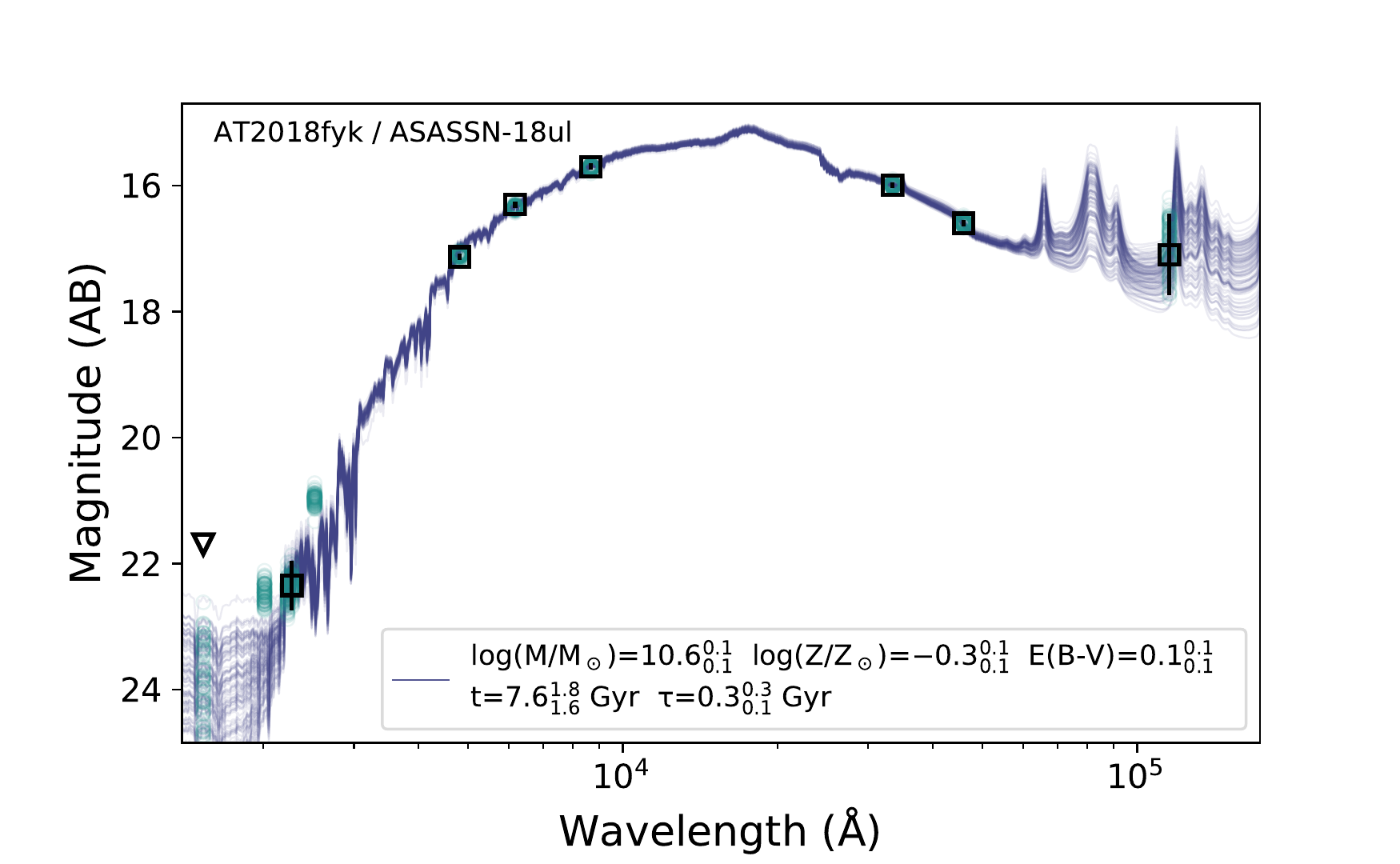}
\caption{ Samples from the posterior distribution of host galaxy SED models. The open squares (and triangle) show detections (and one upper limit) from GALEX, DECaLS, and WISE. Open circles show the synthetic host galaxy magnitude, including predictions for the three bluest bands of the {\it Swift}/UVOT instrument.}
\label{fig:sed}
\end{figure}
To estimate the host galaxy brightness in the {\it Swift}/UVOT bands, we follow the approach of \citet{vanvelzen20}. We use the flexible stellar population synthesis code \citep{Conroy09} and pre-flare host galaxy photometry, including UV (GALEX), optical (DeCALs; \citealt{Dey19}) and IR (WISE; \citealt{Wright10}) observations to model the SED (see Table \ref{tab:hostmags}). The GALEX flux is extracted using the gPhoton software \citep{Million16}, using the same aperture (7") as applied to the UVOT images. For the WISE photometry we adopt the forced photometry values measured by the Legacy survey \citep{Dey19}. 

\begin{table}
    \centering
        \caption{Measurements of the host galaxy brightness in UV, optical and IR filters, used to model the host SED. The model predictions are also presented. Apertures are matched (7 arcsec) for consistency.}
    \begin{tabular}{ccc}
    \hline
Filter & Measured magnitude & Model magnitude\\
& (mag) & (mag) \\
         \hline
GALEX FUV & 23.17$\pm$0.24& 23.26$\pm$0.25\\
GALEX NUV & 22.15$\pm$0.15& 21.78$\pm$0.15\\
DeCALs g & 17.10$\pm$0.04& 16.94$\pm$0.03\\
DeCALs r & 16.34$\pm$0.03& 16.27$\pm$0.03\\
DeCALs z & 15.70$\pm$0.03& 15.66$\pm$0.03\\
WISE W1 & 16.00$\pm$0.03& 16.15$\pm$0.04\\
WISE W2 & 16.61$\pm$0.04& 16.81$\pm$0.04\\
WISE W3 & 16.9$\pm$0.5& 17.0$\pm$0.5\\
Swift UVW2 & ---& 21.98$\pm$0.10\\
Swift UVM2 & ---& 22.00$\pm$0.15\\
Swift UVW1 & ---& 20.56$\pm$0.08\\
Swift U & ---& 18.80$\pm$0.05\\
Swift B & ---& 17.38$\pm$0.04\\
Swift V & ---& 16.60$\pm$0.03\\
\hline
    \end{tabular}
    \label{tab:hostmags}
\end{table}

Markov chain Monte Carlo samples of the posterior distribution are used to constrain the model parameters: stellar mass, stellar population age, metallicity, star formation history e-folding time, and the optical depth of the dust (following the extinction law of \citealt{Calzetti2000a}). The best fit SED model (Figure~\ref{fig:sed}) is used to synthesise host galaxy magnitudes, which are subsequently subtracted from the measured photometry. The uncertainty on the host flux is propagated into the uncertainty on this difference photometry. 

\subsection{UV/optical blackbody analysis}
\label{sec:uvblackbody}
\begin{figure}
\centering
\includegraphics[width=\columnwidth, keepaspectratio]{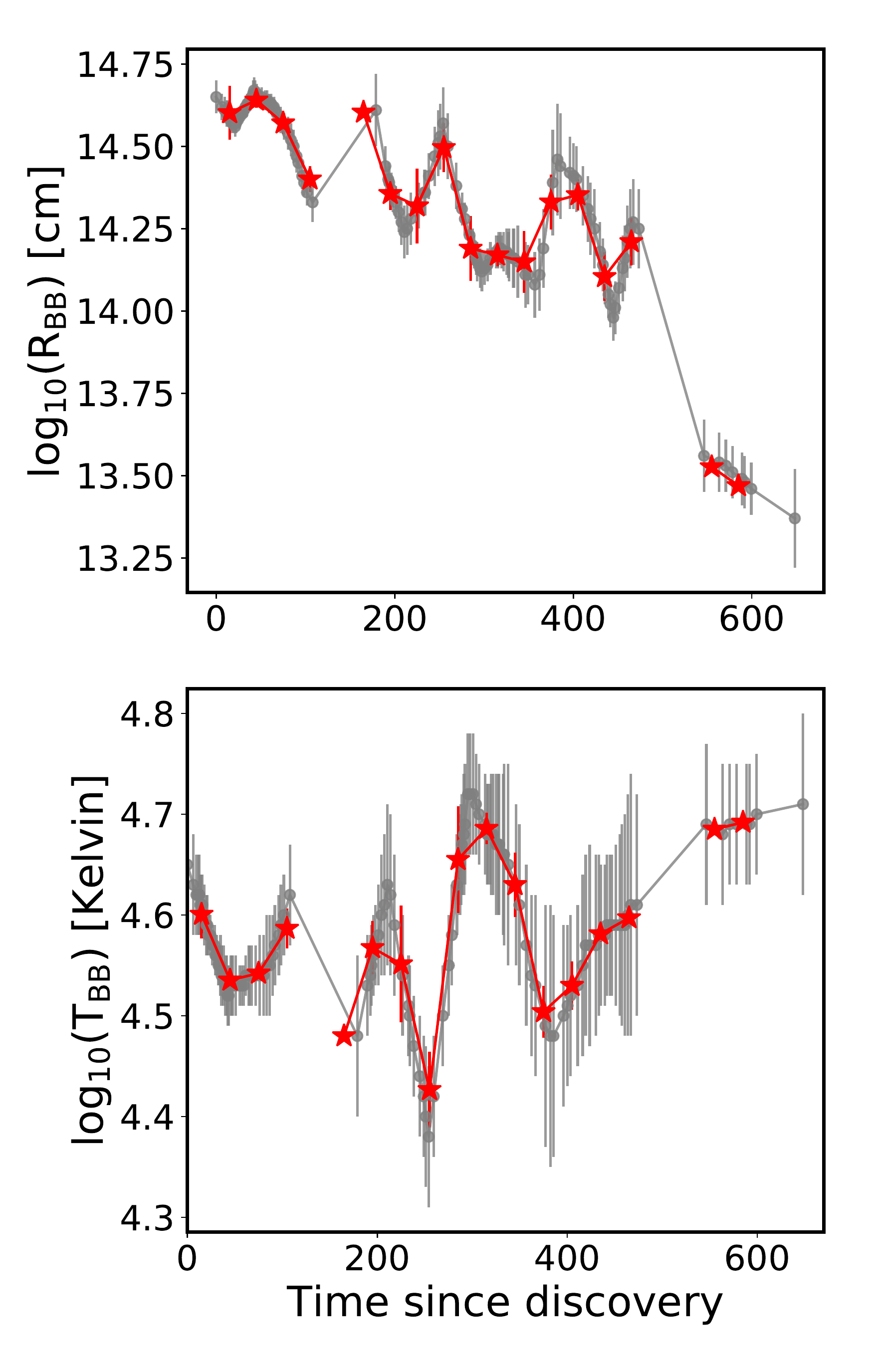}
\caption{ Evolution of the blackbody radius (top panel) and temperature (bottom panel) over time. We show the full time resolution measurements (grey), as well as estimates corresponding to the binned data (red).}
\label{fig:bbrad}
\end{figure}
The UV/optical emission of TDEs can be well described as thermal blackbody emission \citep{vanvelzen11, Hung17}. We therefore measure the blackbody temperature and luminosity by fitting an evolving temperature blackbody model to the light curve. This also yields the blackbody radius, assuming isotropic emission. To smooth out variations induced by poorly constrained measurements, we use a 40 day smoothing length for the temperature measurements, and a 20 day window for the luminosity. Values are linearly interpolated in between grid points.

\begin{figure}
\centering
\includegraphics[width=\columnwidth, keepaspectratio]{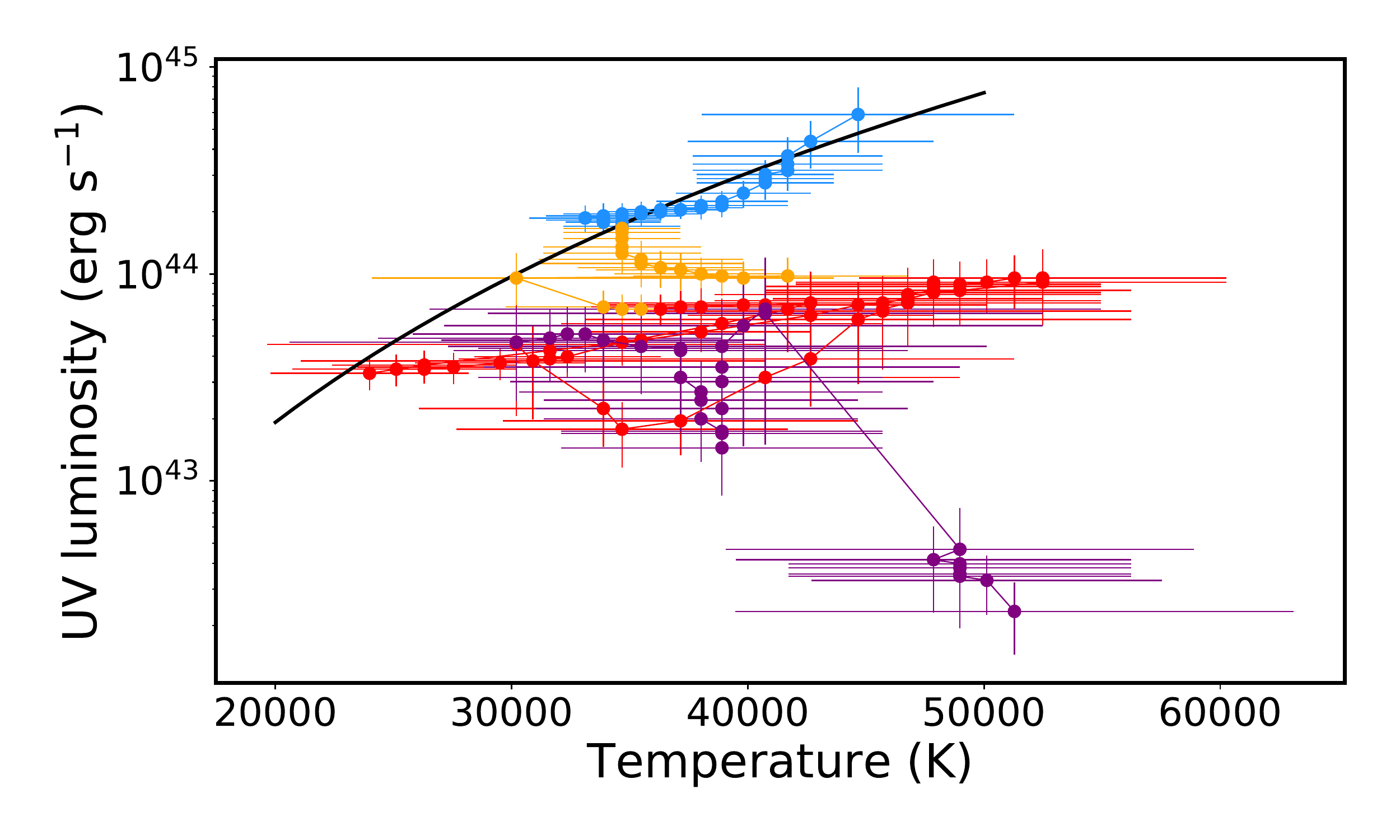}
\caption{TDE UV luminosity as a function of blackbody temperature. The black solid line shows L$\propto$T$^4$ behaviour (with arbitrary normalisation), which roughly describes the behaviour in the soft state (A; blue markers) well. The different states are coloured according to Figure \ref{fig:alphaox}.}
\label{fig:templum}
\end{figure}

The blackbody radius and temperature evolution are shown in Figure \ref{fig:bbrad}. Individual epoch measurements are shown in grey; binned values are shown in red. The long term behaviour of the radius evolves slowly (decreasing only by a factor of $\sim$2) up to 450 days. Similarly, the temperature does not cool significantly over the entire light curve, with variations within the error budget. This is somewhat atypical of other TDEs, although there are sources that show a similar evolution \citep{vanvelzen20}. We do note that during the soft state, there is some temperature evolution consistent with an L $\propto$ T$^4$ behaviour (Figure \ref{fig:templum}). At very late times (560 days after discovery) we find that the radius has decreased significantly, while the temperature (although relatively uncertain) does not show any evidence for significant change.

We interpret this lack of significant temperature and radius evolution during the first 500~d as suggestive evidence that the UV emission originates in a stable, rapidly formed accretion disk. At late times, the luminosity decreases significantly, leading to the inferred decrease in radius.

\subsection{The UV--X-ray spectral slope \alphaox and bolometric Eddington fraction}
\label{sec:alphaox}
\begin{figure*}
\includegraphics[width=\textwidth, keepaspectratio]{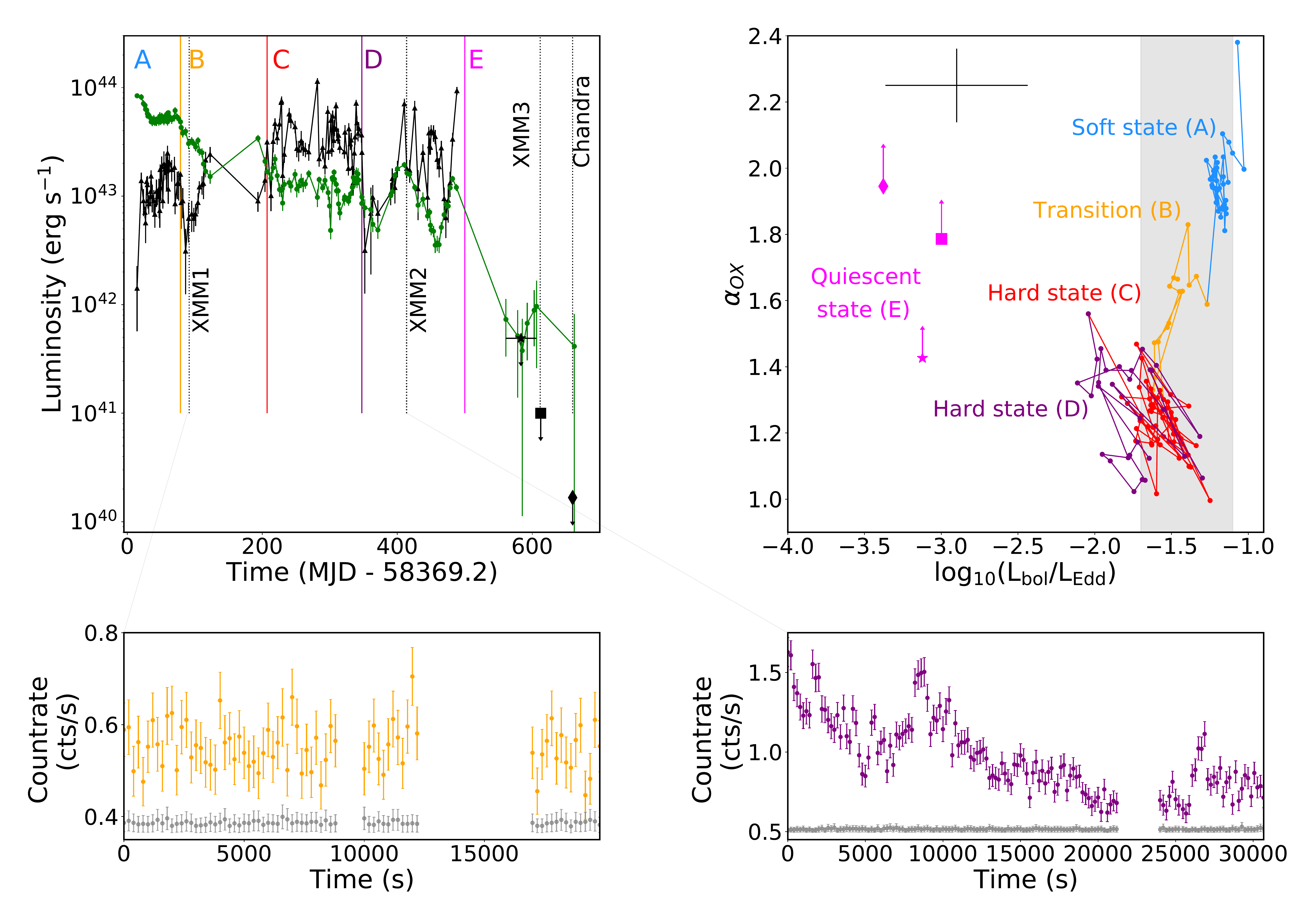}
\caption{Light curves and \alphaox evolution of AT2018fyk. Top left: {\it Swift}/XRT 0.3--10 keV (black triangles) and {\it Swift}/UVW1 (green circles) light curves. The black star, square and diamond represent {\it Swift}, XMM3 and {\it Chandra} X-ray upper limits, respectively. The vertical coloured lines denote transitions in source properties as detailed in Section \ref{sec:alphaox} and Table \ref{tab:phases}, while vertical black dashed lines indicate long stare X-ray observations.
Top right: \alphaox as a function of the bolometric Eddington fraction. A representative error bar is shown in black in the top left; the uncertainty in \boleddfrac is dominated by the intrinsic scatter in the M\,--$\sigma$ relation. The typical transition luminosity of stellar-mass black holes is shown as a grey shaded band. Bottom panels: XMM high cadence (bin size of 200 s) lightcurves during the XMM1 (left) and XMM2 (right) observations. Background rates are shown in grey, offset (for display purposes) by +0.4 and +0.5 for XMM1 and XMM2, respectively.}
\label{fig:alphaox}
\end{figure*}
To probe the broad band long term evolution of the accretion flow, we employ the UV to X-ray spectral index \alphaox \citep{Tananbaum79}:
\begin{equation}
    \alpha_{\rm ox} = 1 - \frac{log_{10}(\lambda L_{2500}) - log_{10}(\lambda L_{2 keV})}{log_{10}(\nu_{2500}) - log_{10}(\nu_{2 keV})}
\end{equation}
where $\lambda$L$_{2500}$ and $\lambda$L$_{2 keV}$ are the monochromatic luminosities at 2500 \AA\ and 2 keV, respectively, and $\nu_{2500}$ and $\nu_{2 keV}$ are the frequencies at those wavelengths. We use the {\it Swift} UVW1 host-subtracted luminosities ($\lambda_{\rm cen}$ = 2629 \AA) as a proxy for L$_{2500}$. These data are included as supplementary data files. Based on the observed behaviour of \alphaox with bolometric Eddington ratio (Figure \ref{fig:alphaox}, top right panel), we divide the TDE evolution into 5 phases (A through E). These phases are defined based on a combination of \alphaox, X-ray spectral and timing properties as well as X-ray and UV variability properties.
We summarise the observational properties for each phase in Table \ref{tab:phases}, but defer a detailed discussion to Section \ref{sec:discussion}.
\begin{table*}
    \centering
        \caption{Observational properties for each phase labelled in Figure \ref{fig:alphaox}. The phase is referenced with respect to the discovery epoch. f$_{\rm Edd}$ is the bolometric Eddington fraction of emission. BB stands for blackbody; PL for power-law.}
    \begin{tabular}{lcccccccc}
    \hline
Phase / State & MJD & Phase & \alphaox & f$_{\rm Edd}$ & X-ray spectral state & X-ray PSD power & UV var. \\
 \hline
A / Soft & 58383 -- 58446 & 14 -- 77 & 2.0 & $\sim$0.1 & BB dominated & Low $\nu$ ($<$10$^{-5}$ Hz) & Yes \\
B / Transition & 58447 -- 58574 & 78 -- 205&1.6 & $\sim$0.05 & --- & Low $\nu$ ($<$10$^{-5}$ Hz) & Yes \\
C / Hard & 58576 -- 58717 & 207 -- 348&1.2 & $\sim$0.03 & PL dominated & Low+high $\nu$ ($<$10$^{-3}$ Hz) & No \\
D / Hard & 58721 -- 58858 & 352 -- 489&1.2 & $\sim$0.03 & PL dominated &Low+high $\nu$ ($<$10$^{-3}$ Hz) & Yes  \\
E / Quiescent & 58930 -- ... & 561 -- ... & $>$1.9 & $<$0.0004 & --- & --- & ---\\
\hline
    \end{tabular}
    \label{tab:phases}
\end{table*}

The UV blackbody temperature does not cool significantly over the 650 day lightcurve, with an average temperature of T$\sim$35000 K.
The UVW1 filter is on the Rayleigh-Jeans tail of the blackbody spectrum with log$_{10}$(T)=4.63 K. Similar to \cite{vanvelzen2019}, we apply a temperature correction to convert the observed luminosity L$_{UVW1}$ to a bolometric UV luminosity estimate L$_{\rm UV}$, integrated from 0.03 to 3 $\mu$m. This correction amounts to a factor of $\sim$6.1, which is adopted for the entire light curve because of the lack of temperature evolution. We further include the integrated X-ray emission in the 0.01--10 keV energy range (including the EUV band) by extrapolating the spectral model from the observed band to define the bolometric Eddington fraction of emission \boleddfrac $\equiv$ (L$_{0.01-10 \rm keV}$ + L$_{\rm UV}$) / L$_{\rm Edd}$. 
A high value of \alphaox indicates a UV (disk) dominated system (also dubbed a soft state), while a low value of \alphaox indicates that the X-ray (corona) emission dominates the energy output (i.e., a hard state).

We use the spectral fits (Section \ref{sec:xrayfitting}) to determine the monochromatic 2 keV X-ray flux L$_{2 keV}$. Given the observed spectral variability with time, we use a spectral model for each of the states A (soft), B (transition) and C+D (hard), and take the best fit power-law index and the appropriate power-law fraction to compute L$_{2 keV}$ for each epoch (Table \ref{tab:xrayspectra}). Because the temperature of the thermal X-ray component is low, the power-law dominates the X-ray flux $>1.5$ keV at all times, and we ignore the thermal contribution in our estimate of L$_{2 keV}$ (i.e. we only consider the power-law component). Uncertainties are propagated using the standard rules.

We note that in the final epoch of {\it Swift} UV observations, the UVM2 brightness is consistent with the SED model prediction for the host galaxy (Table \ref{tab:hostmags}). The UVW1 and UVW2 magnitudes are still slightly elevated when compared to the prediction for the host galaxy. This could suggest that the SED model magnitudes are slight underestimates of the true host brightness. We assess the impact of this difference, in particular for the UVW1 filter, by repeating all calculations while assuming a host galaxy brightness equal to that of the last {\it Swift} UVW1 observation, rather than that of the best fit SED model. The impact on both \alphaox and \boleddfrac is small, with changes of up to $\approx 0.15$ and 0.05, respectively, which doesn't influence the main conclusions of our work.

The left panel of Figure \ref{fig:alphaox} shows the light curves that were used to compute \alphaox and \boleddfrac; note that epochs with only an X-ray or UVW1 measurement are discarded. 
The source spends approximately 65 days in state A, 125 days in state B (although this is an upper limit due to lack of observations in a seasonal gap), and $\sim$280 days in the hard state (C and D) afterwards. State C and D are spectrally identical in X-rays, but the latter is characterised by large amplitude variations in the UV band. At very late times (state E), the UV luminosity decreases by a factor $\sim$15, while the X-ray flux drops by a factor of $>$5000 compared to the last detection with {\it Swift}. 
The right panel shows the corresponding behaviour in \alphaox and \boleddfrac.

\subsection{X-ray spectral analysis}
\label{sec:xrayfitting}
\subsubsection{X-ray spectral state transition}
We employ a two-component phenomenological spectral model to characterise the X-ray spectral evolution following the TDE. We use the 0.3--10 keV band, unless stated otherwise. Our focus is on the evolution of the relative strengths of these two model components during the different phases of the TDE. We start our analysis with the XMM1 and XMM2 EPIC/pn spectra, which contain $\approx$12\,000 and 28\,000 counts for the early and late epochs, respectively. We also tabulate the results of the spectral fitting of the XRT stacked spectra (Table \ref{tab:xrayspectra}) and time-resolved {\it NICER} stacked spectra (0.4-2 keV; Supplementary data file). All analysis is performed using \textsc{xspec} version 12.10.0 in HEASOFT v6.24. Best-fit model parameter uncertainties are estimated using the {\tt error} command in \textsc{xspec}.

We find no evidence for additional intrinsic absorption in the earliest {\it Swift} spectrum, which might be expected if the disk is initially in a slim state \citep{Wen20} (recalling that we find a peak mass accretion rate $\sim$ the Eddington rate). None of the later spectra show evidence for additional neutral absorption either. Therefore, we fix the hydrogen column density to the Galactic foreground value of n$_H$ = 1.15$\times$10$^{20}$ cm$^{-2}$ \citep{hi4pi} for all subsequent analysis. 

For spectral analysis we only used the EPIC-pn data. We bin the two EPIC/pn spectra to a minimum of 25 counts per spectral bin, and oversample by a factor of 3. 
We first attempt to fit the spectra using a single component: a blackbody model ({\tt bbody}), as well as multi-colour blackbody ({\tt diskbb}), power-law ({\tt powerlaw}) and thermally Comptonised continuum models ({\tt Nthcomp}). None of these describe the data well (reduced chi-squared value $\chi_r > 1.6$). Strong systematic residuals are apparent in all cases. 

\begin{figure*}
\includegraphics[width=\textwidth, keepaspectratio]{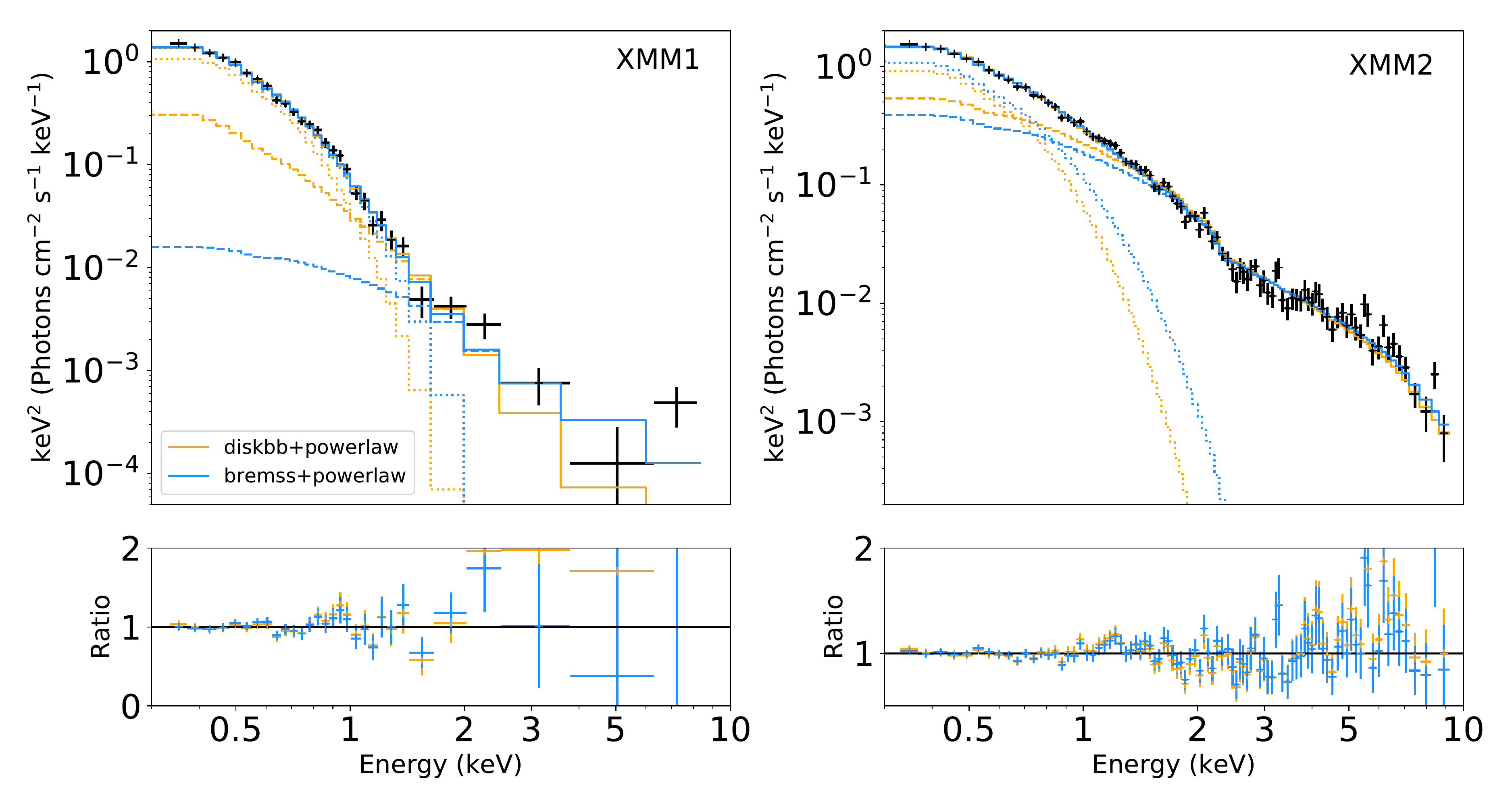}
\caption{XMM-Newton EPIC/pn X-ray spectra and best fit models. {\it XMM-Newton} PN spectra (black), overlaid by two best fit models (diskbb+powerlaw in orange, bremsstrahlung+powerlaw in blue). Dashed lines represent the power-law model, while dotted lines show the thermal component and solid lines are the sum of the two. The bottom panels show the ratio of data and model. The power-law spectral component is stronger in XMM2 than during XMM1.}
\label{fig:xmmspectra}
\end{figure*}

We next employ a thermal model to describe the soft energies ($<$1.5 keV) and a power-law model to describe the higher energies (see Figure \ref{fig:xmmspectra}). Fitting a power-law to higher energies only, we find a very large soft excess below 1.5 keV for XMM1 (factor $>$30 at 0.3 keV), suggesting that relativistic reflection models cannot account for the spectral shape, although we did not explore detailed reflection spectral modelling. We note that the soft excess we refer to is soft X-ray emission in addition to power-law X-ray emission, and does not refer to the pure blackbody spectra observed in some TDEs. We next add a thermal component to describe the soft energies. We first try {\tt TBabs$\times$zashift$\times$(diskbb+powerlaw)} in {\it XSPEC}, which yields temperatures $kT$ = 123 and 146 eV for XMM1 and XMM2, respectively. The power-law index $\Gamma \sim$3.3$\pm$1.0 is steep at early times (although the errors are large), but at late times becomes well constrained $\Gamma$ = 2.1$\pm$0.1. We note that the reduced $\chi^2$ is marginally statistically acceptable for both fits (reduced chi square of $\chi_r = 1.29$ and 1.32, respectively; see Table \ref{tab:xrayspectra}). We find significant evolution in the power-law contribution to the total X-ray flux between XMM1 and XMM2, increasing from $\sim$0.25 to 0.64. Using a more physically motivated model, {\tt TBabs$\times$zashift$\times$(diskbb + nthcomp)}, where the Comptonising seed photon temperature is linked to the accretion disk temperature, yields similar results. 

Employing instead a bremsstrahlung model ({\tt bremss}) for the thermal component provides better quality of fit. Using this model, we find plasma temperatures of $kT$ = 210 and 290 eV, while the power-law indices remains constant at $\Gamma \sim$ 1.9, for $\chi_r$ = 1.08 and 1.07, respectively. In agreement with the {\it diskbb} model results, we find that the power-law fraction of emission increases from near negligible (0.05$\pm$0.03) to providing a dominant (0.54$\pm$0.04) contribution to the total X-ray flux. 

\subsubsection{Power-law fraction of emission}
\begin{figure*}
\includegraphics[width=\textwidth, keepaspectratio]{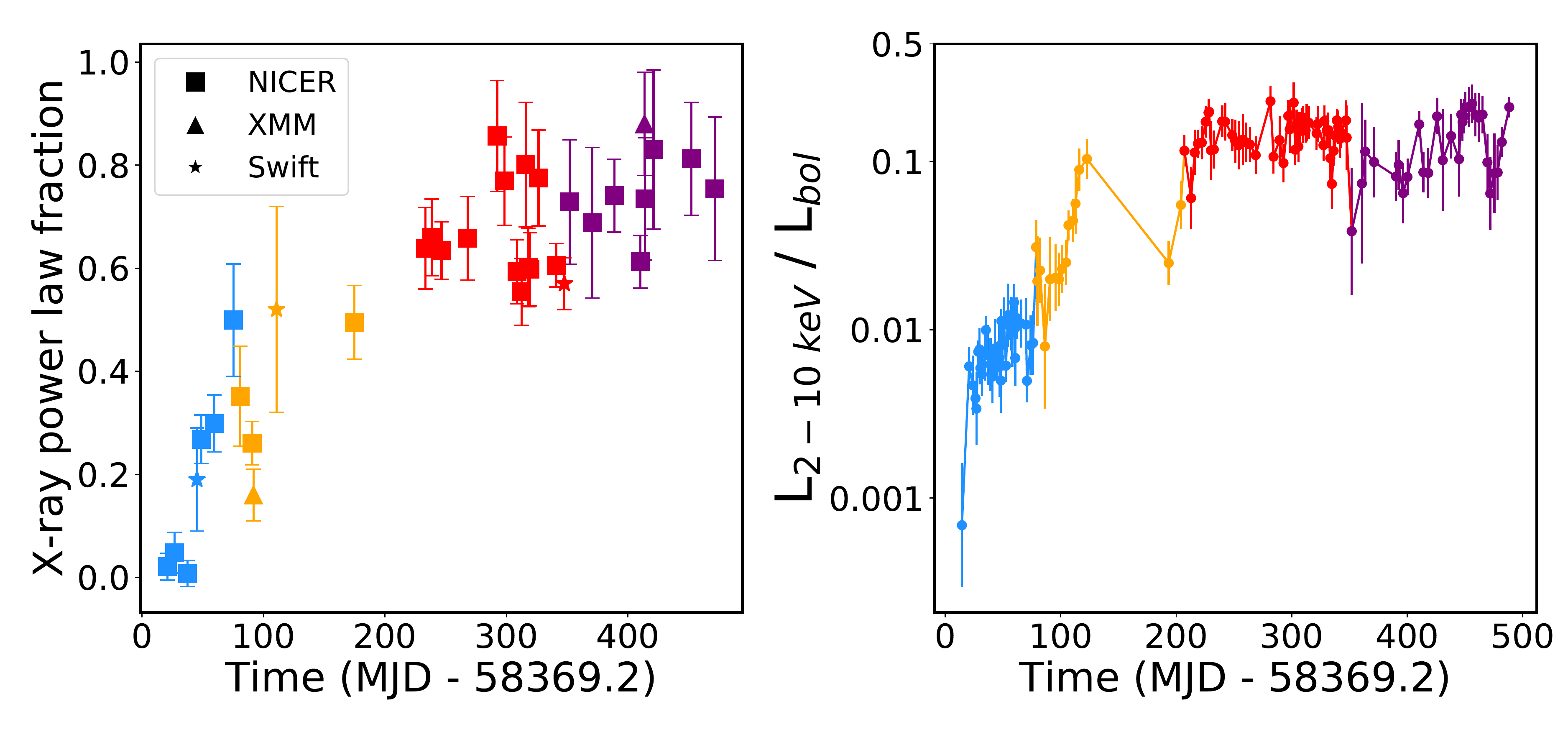}\\
\caption{Long-term evolution of the  X-ray corona. Left: the fractional contribution of the power-law component to the total X-ray flux, i.e., a measure of X-ray corona's strength, derived from {\it NICER} (squares), {\it Swift} (stars) and {\it XMM-Newton} (triangles) spectral fits in the 0.4--2 keV band. Colours are the same as in Figure \ref{fig:alphaox}. The results are consistent with spectral fits in the 0.3--10 keV energy range of {\it Swift} and {\it XMM-Newton} (right panel and Table \ref{tab:xrayspectra}).
Right: fractional contribution of the 2--10 keV power law emission to the total bolometric luminosity, as derived from {\it Swift} measurements. Errors are added in quadrature, but exclude the uncertainty in the M--$\sigma$ relation ($\sim$0.4 dex). The corona contribution is negligible at early times but becomes very significant after +200 days, indicating the strengthening of the corona over time.}
\label{fig:corfrac}
\end{figure*}
Time-resolved spectra from {\it NICER} over the first $\sim$ 450 d provide further clear evidence for significant spectral evolution. Background subtracted, unbinned spectra are fit to the same thermal + power-law models employed for the XMM analysis in {\tt xspec}, using Cash statistics \cite{Cash79}. The fits are conducted in the 0.4--2 keV band, and reported luminosities refer to this energy range. This particular band pass was chosen to minimise the background contamination (see Section \ref{sec:nicerdata} for more details). The effects of imperfect background subtraction can become significant when the power-law flux is low. For power-law fluxes $<$10$^{-13}$ erg cm$^{-2}$ s$^{-1}$, the power-law index is therefore fixed at the best fit value of the composite ($\sim$310 ks) spectrum to mitigate these effects.

\begin{figure}
\includegraphics[width=\linewidth, keepaspectratio]{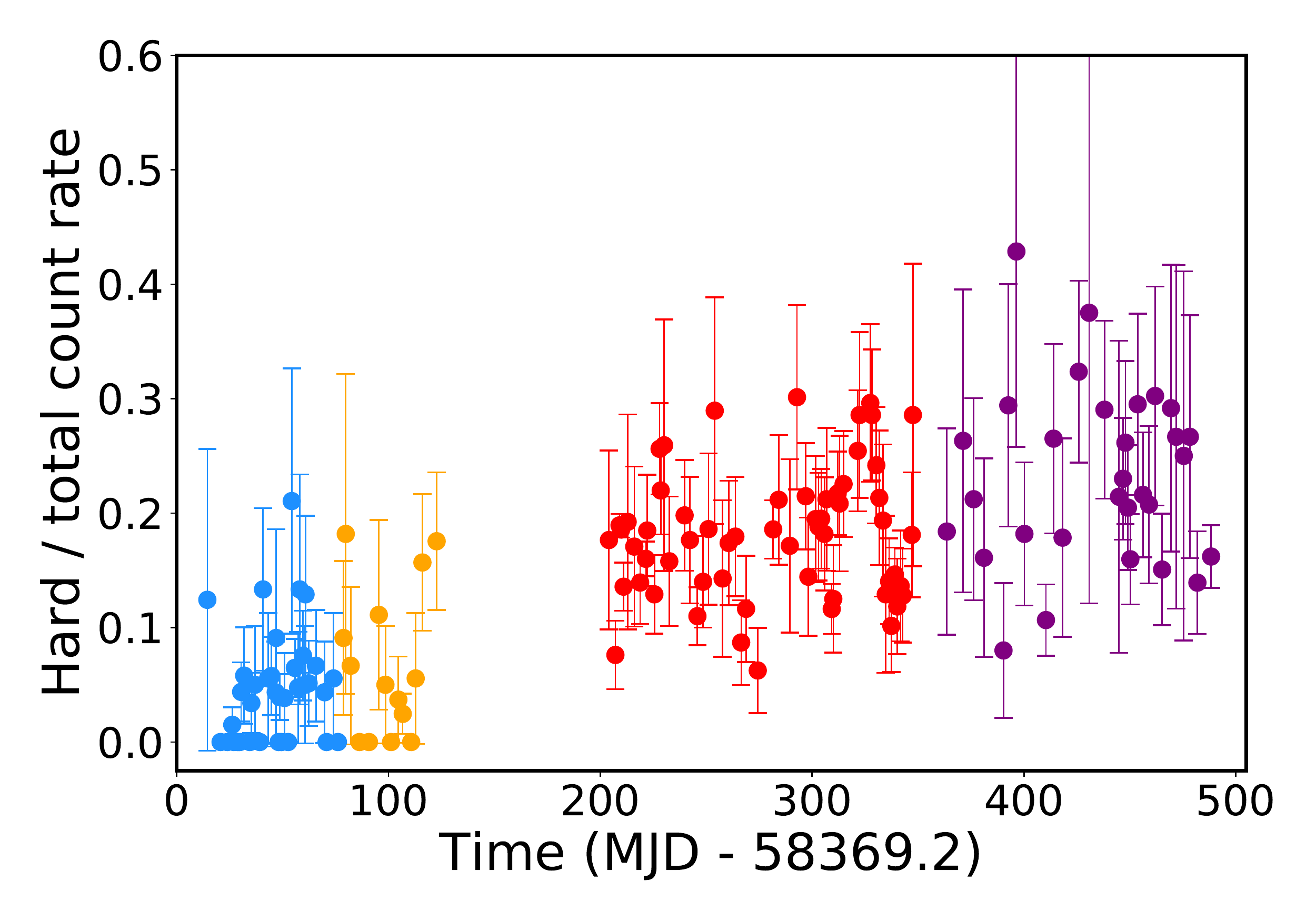}
\includegraphics[width=\linewidth, keepaspectratio]{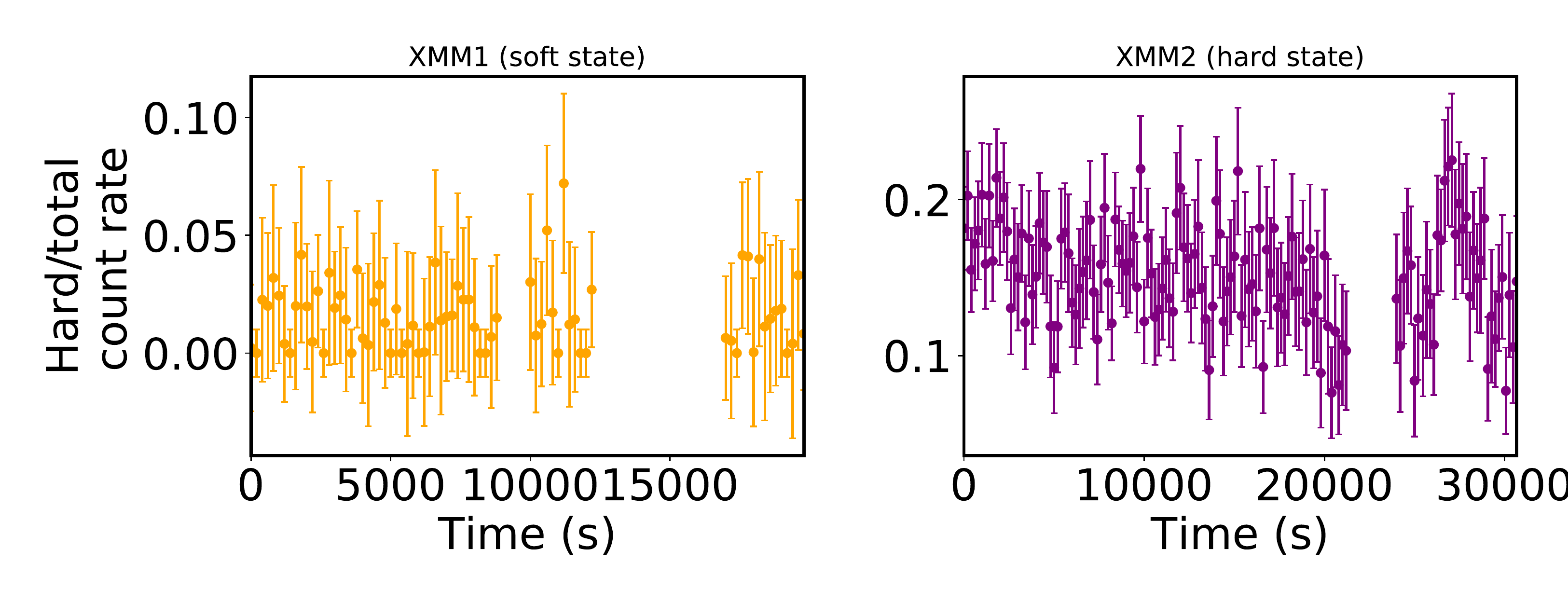}
\caption{Hard to total count rate ([1.5--10 keV]/[0.3--10 keV]) as a function of time as measured by {\it Swift}/XRT (top) and XMM-Newton/EPIC (bottom). The evolution is consistent with results from time-resolved spectra shown in Figure \ref{fig:corfrac}, i.e., hard X-rays dominate in states C and D.}
\label{fig:swift_hr}
\end{figure}

The fractional contribution of the power-law component with respect to the total flux is estimated from each of these fits. The evolution of the power-law fraction of X-ray emission, i.e. the relative strength of the corona, are shown in Figure \ref{fig:corfrac} (left panel) for the {\tt diskbb+powerlaw} model. For consistency, we also refit the {\it Swift} and {\it XMM-Newton} spectra in the 0.4--2 keV energy range; we note that these results are consistent with those obtained from 0.3--10 keV spectral fits for {\it Swift} and {\it XMM-Newton} observations. The spectra at early times (state A and early state B) are dominated by a thermal component, which provides 50-–95 \% of the X-ray flux (0.3--10 keV), indicating the presence of a weak corona. In the hard states the corona strengthens to provide $\sim 60-80$ per cent of the X-ray flux. A similar increase in power-law fraction from near negligible ($<15$ per cent) in the soft state to very significant (20--75 per cent) in the hard state is also found for the {\tt bremss+powerlaw} model. To remain agnostic about the nature of the soft excess emission, we perform a similar calculation but this time taking the ratio of power-law flux over the bolometric emission (Figure \ref{fig:corfrac} right panel). This yields a qualitatively very similar picture, with an initially weak corona that subsequently strengthens over time. This evolution is further corroborated by the {\it Swift} and {\it XMM-Newton} data (Figure \ref{fig:swift_hr}), which shows a steady increase in the fraction of hard X-ray count rate to total X-ray emission. \\

We conclude that there is significant spectral evolution in AT2018fyk, where the corona is energetically unimportant at early times, while it produces more than half of the observed X-ray flux, and a significant fraction of the bolometric emission, at late times. This evolution is consistent with a spectral state transition from a soft state to a hard state, analogous to those seen in XRBs, AGN \citep{Done05, Remillard06} and a sample of X-ray bright TDEs \citep{Wevers20}.

\subsubsection{Rapid spectral variability}
In terms of rapid ($\sim$minutes to tens of hours timescale) spectral variability, the hardness ratio of the X-ray emission as observed by {\it Swift} and {\it XMM-Newton} can vary on short timescales (Figure \ref{fig:swift_hr}). However, this variability does not obviously correlate with the X-ray flares. 
We do not have high quality spectroscopic X-ray data to assess spectral variability on rapid timescales of tens of minutes (e.g. as seen in quasi-periodic eruptions \citealt{Miniutti19, Giustini20}). Analysis of {\it Swift} spectra stacked based on their X-ray count rate does not reveal significant spectral changes.
The {\it XMM-Newton} hardness fraction is constant during the soft state (XMM1) with an average value of 0.016$\pm$0.015, i.e. consistent with 0. However, the hardness fraction varies smoothly between 0.07 and 0.22, with an average value of 0.15$\pm$0.03 during the hard state (XMM2). Interestingly, roughly 27000 s after the start of the exposure there is a hardness flare that coincides with a flux enhancement in the light curve (see Figures \ref{fig:alphaox} and \ref{fig:swift_hr}). 

\begin{deluxetable*}{cccccccccccc}
\tablecaption{Best fit parameters obtained from X-ray spectral modelling of {\it Swift} and {\it XMM-Newton} data. The states are as labelled in Figure \ref{fig:alphaox}; states C and D are spectrally identical, and only 1 Swift stacked spectrum is produced covering both. The mean count rate for each spectrum is given in the second column. The effective exposure time t$_{\rm exp}$ is given in kiloseconds. The spectral model used is {\tt TBabs$\times$zashift$\times$(model + powerlaw)}, where model is listed in the table; kT is the temperature of the thermal component, while $\Gamma$ denotes the power-law spectral index. Normalisations for the thermal and power-law components are listed in the norm(kT) and norm($\Gamma$) columns. The flux is integrated from 0.3--10 keV. PL frac denotes the fractional contribution of the power-law component to the total X-ray flux. The final column lists the reduced $\chi^2$ and degrees of freedom (dof).\label{tab:xrayspectra}}
\tablewidth{700pt}
\tabletypesize{\scriptsize}
\tablehead{
\colhead{Spectrum} & \colhead{Count rate} & \colhead{State} & \colhead{t$_{\rm exp}$} & 
\colhead{Model} & \colhead{kT} & \colhead{norm(kT)} & \colhead{$\Gamma$} & 
\colhead{log$_{10}$(norm ($\Gamma$))} & \colhead{log$_{10}$(flux)} & \colhead{PL frac} & \colhead{$\chi^2$ (dof)} \\ 
\colhead{} & \colhead{(cts s$^{-1}$)} & \colhead{} & \colhead{(ks)} & 
\colhead{} & \colhead{(eV)} & \colhead{} &
\colhead{} & \colhead{} & \colhead{(erg cm$^{-2}$ s$^{-1}$)} & \colhead{}
} 
\startdata
         XRT & 0.037 &  A & 37.6 & diskbb & 142$\pm 10$ & 260$\pm$150& 2.8$^{+1.1}_{-1.0}$ & --4.2$\pm$0.3 & --11.89$\pm$0.03 & 0.25$\pm 0.13$ & 1.30 (40) \\\vspace{1.5mm}
          &  & &  & bremss & 260$\pm 25$ & 0.0037$\pm$0.0005 & 1.8$^{+1.4}_{-1.1}$ & --4.7$\pm$0.5& --11.87$\pm 0.03$ & 0.10$^{+0.12}_{-0.04}$ & 1.24 (40) \\\vspace{1.5mm}
         EPIC/pn & 0.87 & B & 13.8 & diskbb & 123$^{+8}_{-4}$ & 506$^{+103}_{-238}$ & 3.3$^{+1.1}_{-1.0}$ & --4.5$^{+0.2}_{-0.3}$ & --11.96$\pm$0.02 & 0.27$^{+0.26}_{-0.20}$ & 1.29 (26) \\\vspace{1.5mm}
          & & &  & bremss & 210$\pm 8$ & 0.0050$\pm$0.0004 & 1.8$^{+0.8}_{-0.7}$ & --5.0$\pm$0.3& --11.94$\pm 0.01$ & 0.05$\pm 0.03$ & 1.08 (26) \\\vspace{1.5mm}
         XRT & 0.097 & B & 30.0 & diskbb &166$\pm 15$ & 160$^{+80}_{-50}$ & 2.11$\pm 0.17$ & --3.37$\pm$0.06& --11.44$\pm 0.02$ & 0.63$\pm 0.07$ & 0.82 (95) \\\vspace{1.5mm} 
          & & &  & bremss &330$\pm 50$ & 0.0034$\pm$0.0006& 2.0$\pm 0.2$ &--3.43$\pm$0.03 & --11.43$\pm 0.02$ & 0.43$^{+0.05}_{-0.10}$ & 0.86 (95) \\\vspace{1.5mm} 
         XRT & 0.13 & C+D & 110.4 & diskbb &179$\pm 8$ & 131$\pm$25 & 2.19$\pm 0.08$ & --3.19$\pm$0.05 & --11.31$\pm 0.01$ & 0.68$\pm 0.04$ & 0.92 (214) \\\vspace{1.5mm}
          & & &  & bremss &372$\pm 26$ & 0.0036$\pm$0.0003 & 2.06$\pm 0.11$ & --3.27$\pm$0.05 & --11.31$\pm 0.01$ & 0.59$\pm$0.05 & 0.98 (214) \\\vspace{1.5mm}
         EPIC/pn & 1.13 & D & 24.6 & diskbb & 146$\pm 7$ & 171$^{+40}_{-30}$ & 2.10$\pm 0.08$ & --3.60$\pm$0.03 & --11.69$\pm 0.01$ & 0.64$\pm 0.03$ & 1.32 (90) \\\vspace{1.5mm}
          & &  &  & bremss & 292$\pm 19$ & 0.0024$\pm$0.0002 & 1.93$\pm 0.09$ & --3.69$\pm$0.05 &  --11.67$\pm 0.02$ & 0.54$\pm 0.04$ & 1.07 (90) \\
\enddata
\end{deluxetable*}

\subsection{X-ray timing analysis}
\label{sec:xraytiming}
\begin{figure*}
\includegraphics[width=0.5\textwidth, keepaspectratio]{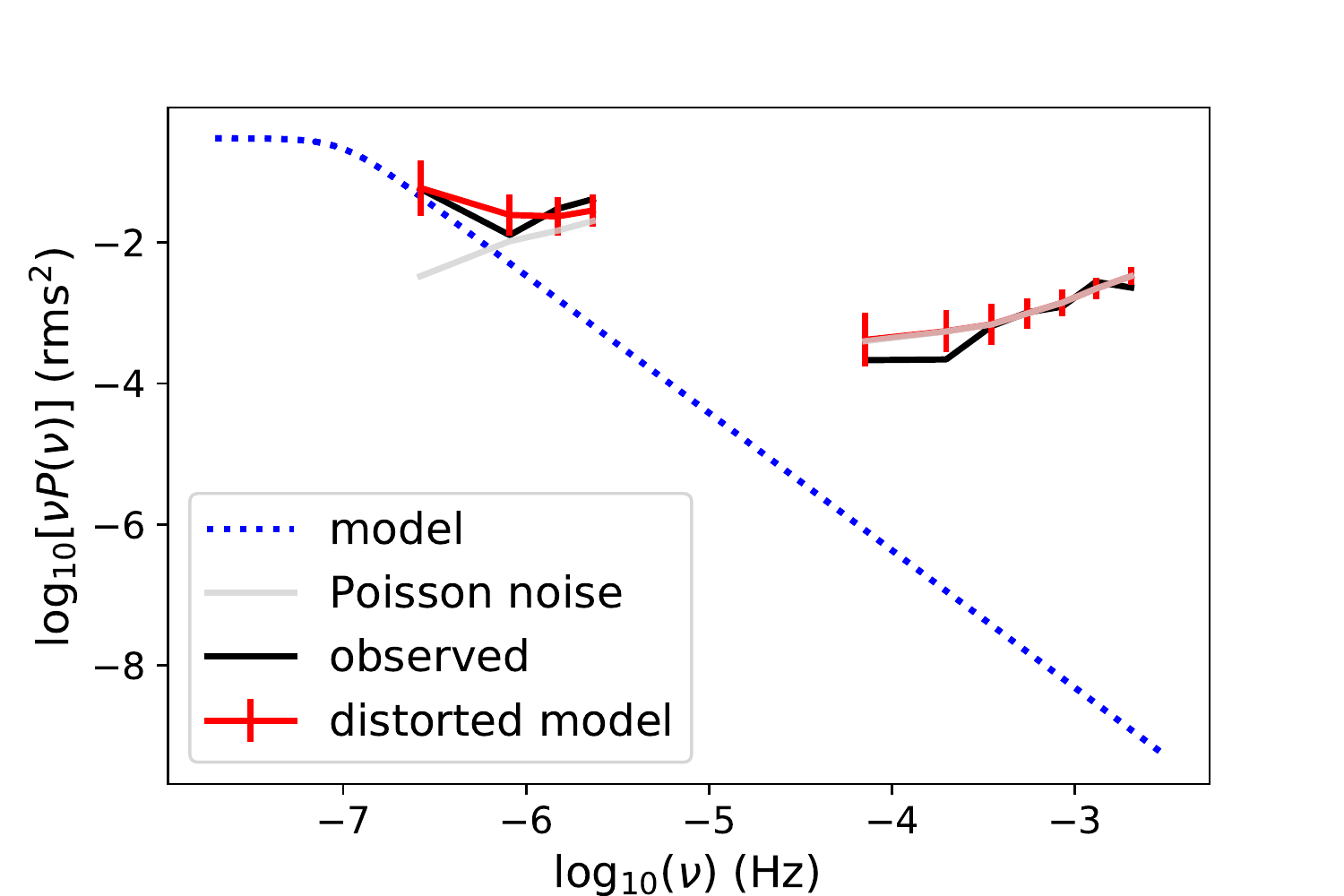}
\includegraphics[width=0.475\textwidth, keepaspectratio]{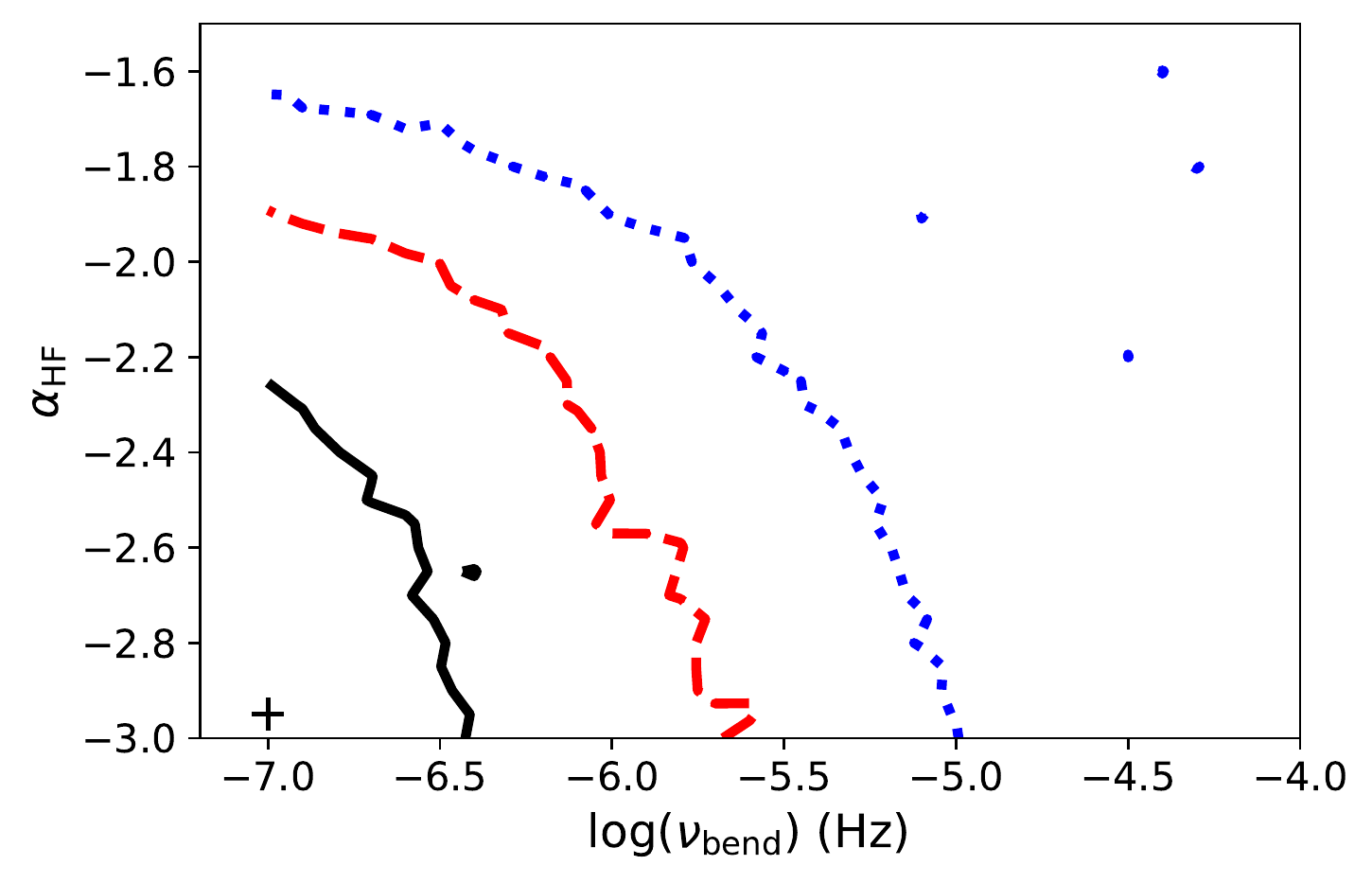}
\includegraphics[width=0.5\textwidth, keepaspectratio]{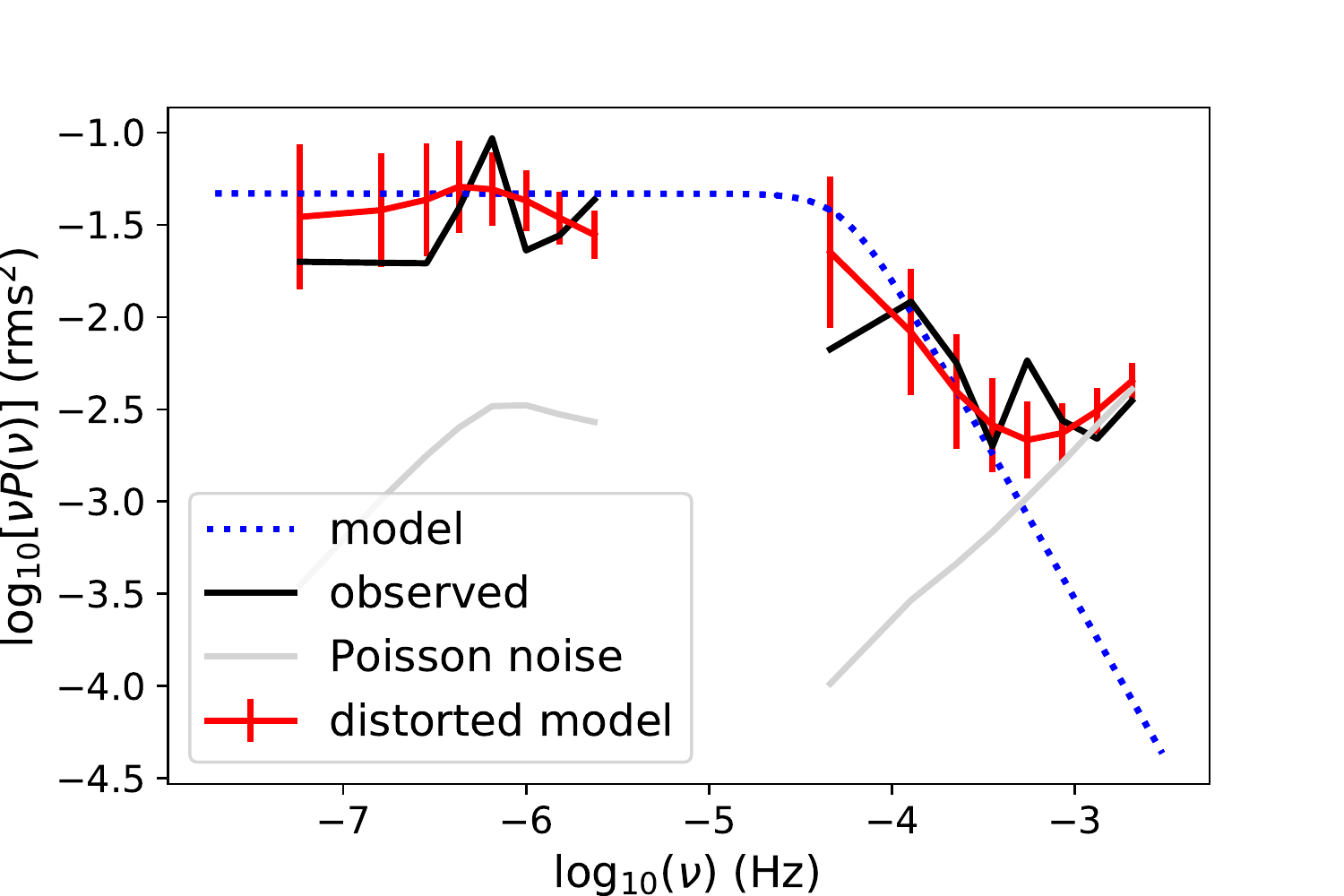}
\includegraphics[width=0.46\textwidth, keepaspectratio]{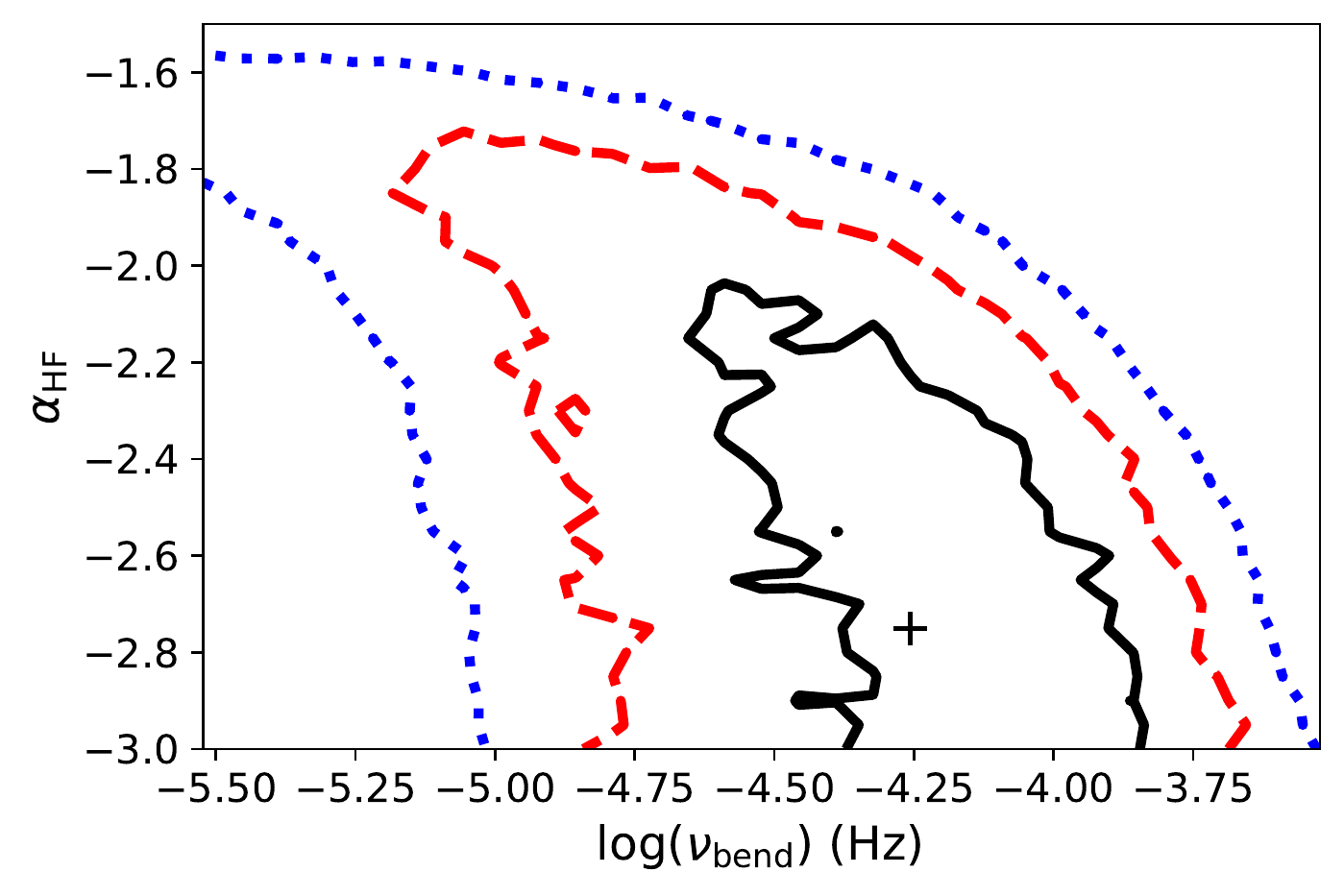}
\includegraphics[width=0.5\textwidth, keepaspectratio]{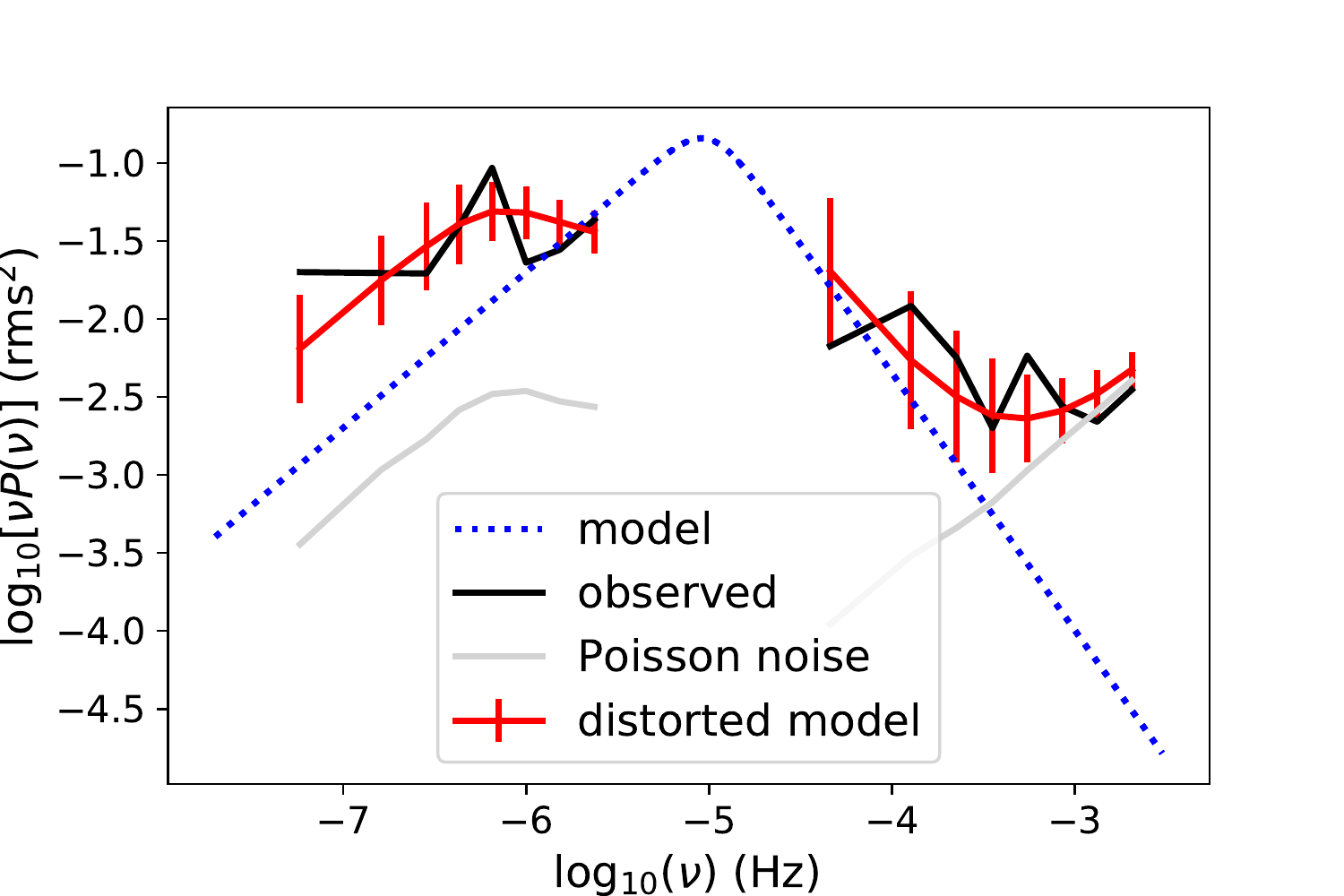}
\includegraphics[width=0.49\textwidth, keepaspectratio]{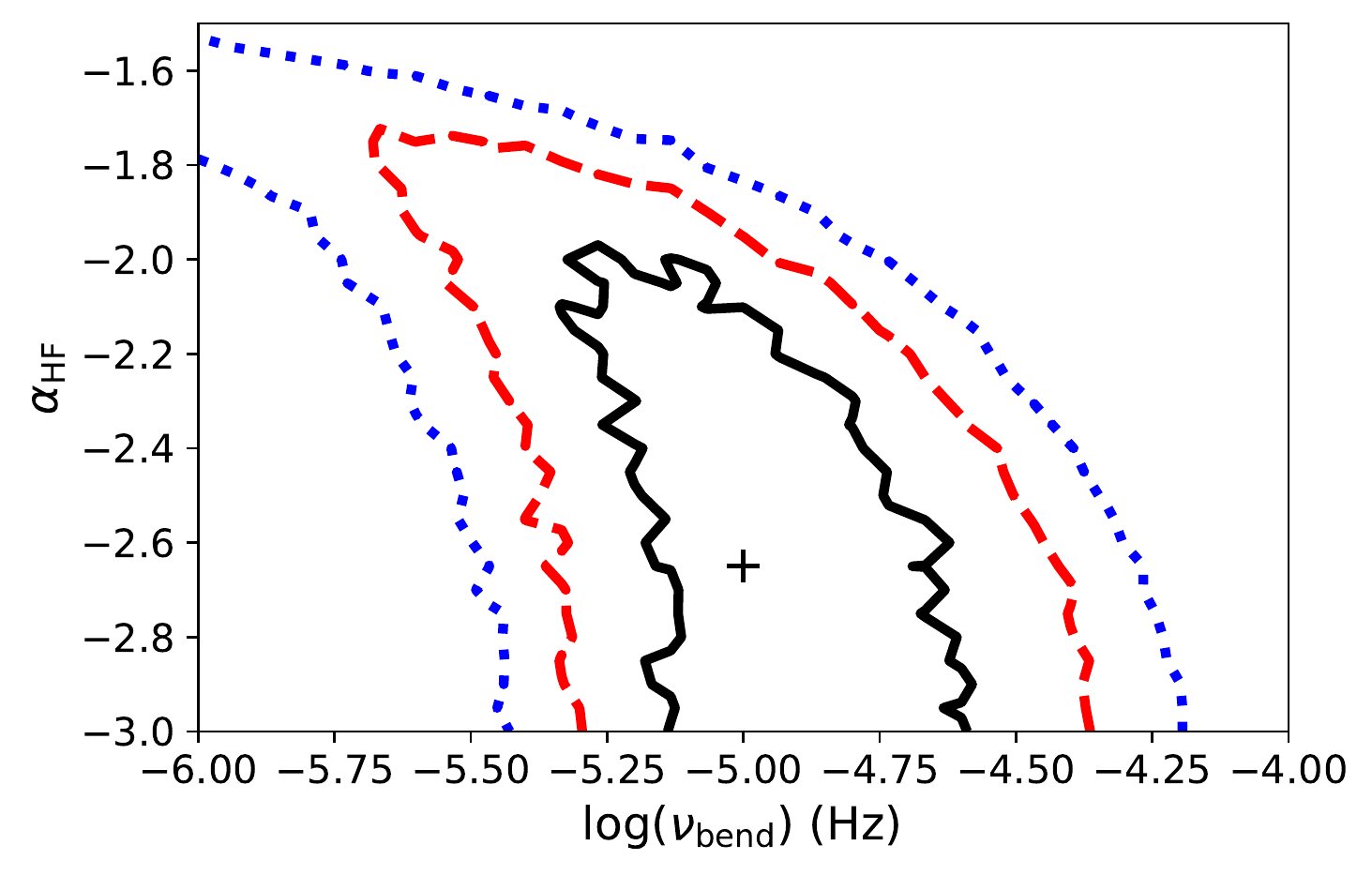}\\
\\
\caption{Power spectra and confidence contours in the soft (top panels) and hard states (middle/lower panels). Left: power density spectra. Observations are shown as solid black lines; the best fit distorted model in red; the undistorted model in blue; and Poisson noise levels are indicated in grey. The soft state is shown in the upper panel; no bend in the power-law model is required to reproduce the data. For high frequencies ($\gtrsim$10$^{-6}$ Hz) the data (black) is consistent with Poisson noise (grey curve). The middle and lower panels shows the hard state, with the low frequency slope fixed to $\alpha_{LF}$ = --1 and $\alpha_{LF}$ = 0, respectively. Variability on frequencies between 10$^{-7}$ to 10$^{-3}$ Hz is evident; a bend is required to reproduce the data. Right panels: corresponding confidence contours between the bend frequency and the high-frequency slope of the power spectrum for 1, 2 and 3-sigma confidence regions (solid black, red dashed, blue dotted lines), corresponding to $\delta \chi^{2} \simeq 2.3$, 6.18 and 11.83 respectively. Black crosses mark the best fit set of parameters.}
\label{fig:psresp_comparison}
\end{figure*}

To constrain the power spectra for the intervals of the light curve identified with the soft (state A) and hard (state C+D) states (including the corresponding {\it XMM-Newton} light curves) we fit the power spectra using the power-spectral response (PSRESP) method \citep{Uttley02}. PSRESP compares the observed power spectra with those of simulated light curves for a given power-spectral shape, resampled to match the gaps in the sampling of the real light curves, so that the distorting effects of sampling on the power spectrum are properly accounted for.

For the {\it Swift}/XRT light curves we first rebin the light curves to give one flux measurement per observation and remove a small amount of data with signal to noise (S/N) $<$1. The light curves are then further rebinned into 2-day bins (slightly larger than the average sampling interval of 1.8 d in states C+D and 1.2 d in state A) and gaps are filled by linear interpolation. The {\it XMM-Newton} light curves are binned at 200 s resolution and gaps are also filled by linear interpolation. We calculate the observed power spectra from the resulting light curves, binning the log-power values geometrically in frequency by a factor of 1.5 (with a minimum of 2 data points per frequency bin).

The Monte Carlo simulated underlying time-series are generated using a sharply bending power-law model, parameterised by low frequency index $\alpha_{\rm LF}$, high frequency index $\alpha_{\rm HF}$ and bend frequency $\nu_{\rm bend}$. For the {\it Swift}/XRT data the simulation time-resolution is set to 5760 s to roughly match the Swift orbital period and ensure that there are 30 simulated data points per 2-day bin. We account for the missing simulated power on time-scales from 5670 s down to the duration of each Swift snapshot ($\sim$1 ks) by calculating the rms of the model on those time-scales and adding a Gaussian random number with the same standard deviation to each sampled simulated data point. By accounting for random variations, this is an advance over the original PSRESP, which accounted for this effect by adding a constant component to the simulated power spectrum and assuming that the log-power errors scaled with those of the simulation.

To match the observed data, 200s sampling is used for the {\it XMM-Newton} light curve simulations. In the PSRESP software, the simulations are resampled to match the sampling of the unbinned light curves, and rebinned and interpolated in the same way as the data, so the effects on the power spectrum are accounted for. 

For each observed light curve, we generate 1024 simulated light curves as segments cut from a single very long light curve (to account for low frequency 'red-noise leak', \citealt{Uttley02}). Unlike in the original PSRESP, which added observational ('Poisson') noise as a constant power spectral component, we include the effects of observational noise by simulating separate noise data sets from Gaussian errors which match the observed errors on each data point, resampling, interpolating and adding the resulting noise power spectrum to the simulated light curve power spectrum. By separating the noise contribution in this way, we can use an arbitrary normalisation to fit the simulated intrinsic power spectra to the data, without needing to specify the observed S/N as an additional variable to be stepped through in the parameter search.

As in the original PSRESP, the simulated light curves are used to calculate the mean and standard deviation on the log-power in each frequency bin, and for a given set of model parameters we can compare this with the observed power spectrum to yield a `pseudo' $\chi^{2}$ value for the distorted power spectrum, $\chi^{2}_{\rm dist}$ of a given pair (Swift/XRT and XMM-Newton) of light curves. We minimise $\chi^{2}_{\rm dist}$ to obtain the best fit normalisation for the given shape parameters of the power-spectral model (power-law indices and bend frequency). By comparing the $\chi^{2}_{\rm dist}$ of the observed power spectrum with those of the simulations for a given model, we can estimate a goodness-of-fit for that model.

In the original PSRESP, contours of goodness-of-fit are used to obtain errors on the power-spectral parameters. However as noted by \cite{Marshall15}, this approach can lead to acceptance regions which are too small when the best fit goodness-of-fit is itself low. From the comparison of $\chi^{2}_{\rm dist}$ of the best fit models with the goodness-of-fit derived from the simulations, we are able to identify that for this particular set of light curve samples, $\chi^{2}_{\rm dist}$ behaves close to an undistorted $\chi^{2}$ distribution. Therefore, we estimate confidence contours on the model parameters using the standard $\Delta \chi^{2}$ approach (which is equivalent to likelihood ratio, for normally distributed errors).

To limit the parameter space covered, we consider two `typical' values of low frequency slope, $\alpha_{\rm LF}=0$ (corresponding to the low frequency shape of the `band-limited noise' seen in black hole XRB hard states) and $\alpha_{\rm LF} = -1$, corresponding to the intermediate frequency slope of `broadband noise' seen in faint black hole XRB hard states, or the low frequency slope that is seen in some soft states seen of black hole XRBs (e.g. see \citealt{Heil15} for a comparison of the range of XRB power-spectral shapes). For each of these two low frequency slopes, we search a parameter space covering a wide range of bend frequency (searched in equally-spaced intervals of log(bend-frequency) and high frequency slope (searched in equal intervals of the power-law index). We obtain confidence intervals on the PDS by Monte Carlo likelihood sampling of the fitted model parameters, the bend frequency and the high frequency slope.

Both low frequency shapes, $\alpha_{\rm LF}=0$ and $\alpha_{\rm LF} = -1$, provide adequate fits to the hard state data over some of their parameter space, with best fit goodness-of-fits of 0.28 and 0.19 respectively. The confidence contours for the two types of model considered and plots of the best fit model (distorted and underlying) versus the data, are shown in Figure \ref{fig:psresp_comparison}. The best fit parameters, their 1-dimensional (1-$\sigma$) errors and fractional rms (integrated from the model over $10^{-7}$--$10^{-3}$ Hz) are provided in Table \ref{tab:psresp}. The high frequency slopes we measure are similar to those measured for AGN samples, which have slopes of $\alpha_{\rm HF} \approx -2$ (if no bend is required for the PDS) to --3 (if the PDS requires a bending power-law; \citealt{Gonzalez12}). Note that the high frequency slope is not well constrained for large negative indices due to red-noise leak effects \citep{Uttley02}.

Since either model provides a good fit to the hard state interval, we cannot distinguish the type of hard state solely on the basis of the power-spectral shape, but we note that the inferred broadband fractional rms of 55~per~cent is much more consistent with black hole X-ray binary hard states than with the soft states (which show rms $<$a few per cent). 

\begin{deluxetable*}{cccccccc}
\tablecaption{Results of the power-spectral fitting. The rms values are reported in various frequency bands. The black hole mass is calculated using the bend frequency in combination with the relation of \cite{Mchardy06}. Values marked with an asterisk have negative error bars bounded by the parameter range. $^{\dagger}$99\% confidence upper limit. ${^\ddagger}$99\% confidence lower limit (the upper range of rms is not well constrained since the low frequency power-spectral shape is not well constrained by the soft state light curve).}
\label{tab:psresp}
\tablewidth{700pt}
\tablehead{
\colhead{State} & \colhead{log($\nu_{\rm bend}$)} & \colhead{$\alpha_{\rm HF}$} & \colhead{rms (\%)} & 
\colhead{rms (\%)} & \colhead{rms (\%)} & \colhead{rms (\%)} & \colhead{log(M$_{BH}$)} \\
\colhead{} & \colhead{(Hz)} & \colhead{(\footnotesize{10$^{-7}$--10$^{-3}$ Hz})} & \colhead{(\footnotesize{10$^{-7}$--10$^{-6}$ Hz})} & 
\colhead{(\footnotesize{10$^{-6}$--10$^{-5}$ Hz})} & \colhead{(\footnotesize{10$^{-5}$--10$^{-3}$ Hz})} & \colhead{(M$_{\odot}$)}
} 
\startdata
Hard ($\alpha_{\rm LF}=0$) & --5.00$_{-0.27}^{+0.33}$  & --2.65 + 0.5$^{*}$ & 54$\pm$1 & 13$_{-3}^{+4}$ & 39$^{+5}_{-8}$ & 35$^{+7}_{-8}$  & 7.25$\pm$0.55\\
Hard ($\alpha_{\rm LF}=-1$) & --4.26$_{-0.20}^{+0.33}$ & --2.75 + 0.5$^{*}$  & 55$_{-3}^{+1}$ & 32$^{+4}_{-3}$ & 32$^{+5}_{-8}$ & 31$^{+7}_{-8}$  & 6.9$\pm$0.55\\
Soft & $<$ --5.5$^{\dagger}$ & --2.95 + 0.5$^{*}$ & 21$^{\ddagger}$ & 32$_{-7}^{+4}$ & 7$^{+2}_{-1}$ & 1$^{+1}_{-0.5}$  & ---\\\hline
\enddata
\end{deluxetable*}

\section{Discussion}
\label{sec:discussion}
The emission of the outer accretion disk around a 10$^{7.7}$ M$_{\odot}$ black hole peaks in the UV band, which we monitored using the Neil Gehrels {\it Swift} satellite. {\it Swift} also performed monitoring observations in X-rays (0.3--10 keV), which contains information about the poorly understood components referred to as the soft excess (dominating the 0.3--1.5 keV band) and the X-ray corona (emitting primarily between 2 and 10 keV). The nature of the former component is still unclear, but it could originate in the inner accretion disk either directly (although the observed temperature may be too high for this in AT2018fyk) or from a low temperature Comptonisation component in the inner regions, while the standard corona likely originates in a hot plasma above the inner disk. We also monitored for radio emission to search for signatures of a newly launched jet. 
These data provide three key diagnostics to characterise the accretion properties following the TDE, which together enable detailed comparison with stellar-mass black holes.

First, the long term light curves (top-left panel of Figure \ref{fig:alphaox}) allow us to track the evolution of the spectral energy distribution (SED), using the UV to X-ray spectral slope \alphaox (Section \ref{sec:alphaox}). This quantity represents the relative strength of the accretion disk (in the UV at 2500 \AA) and the corona (at 2 keV), and serves as a proxy for accretion state. In stellar-mass black holes, the strength of the disk relative to the corona changes as a function of the Eddington ratio \boleddfrac \citep{Fender04, Remillard06}. Appropriate mass scaling of the relative component strengths, represented by \alphaox in SMBHs, also reproduces the observed properties of large samples of AGN and relatively fast evolving changing-look AGN \citep{Sobolewska11a,Ruan19a}; we reiterate that there is potentially a significant (systematic) uncertainty in the black hole mass estimate for AT2018fyk, which should be taken into account when making the comparison to (CL-)AGN behaviour. Given that the uncertainty is systematic, the shape of the evolution throughout the diagram will not change as a result.

Second, to characterise the X-ray spectra in the different phases, we use a two component spectral model (Section \ref{sec:xrayfitting}), including a thermal component to describe the soft energies (0.3--2 keV) and a power-law to account for the harder X-rays (2--10 keV). The key diagnostics are the ratio of the power-law spectral component flux to the total emission in X-rays, and the ratio of power-law emission to the total bolometric luminosity (including the accretion disk, soft excess and corona). These quantities are indicators of accretion state, and their evolution is presented in Figure \ref{fig:corfrac}. 

Third, photometric variability properties of the X-ray emission are analysed by using power density spectra (PDS) derived from the {\it Swift} and {\it XMM-Newton} light curves in the different phases (Figure \ref{fig:psresp_comparison}, see also Section \ref{sec:xraytiming}). The {\it Swift} light curves provide constraints on long timescales (1--25 days), while the {\it XMM-Newton} light curves probe short timescales (minutes -- hours). 
In Figure \ref{fig:psresp_comparison} we compare the soft and hard spectral state power spectra. The best fit model parameters are constrained using Monte Carlo light curve simulations, which are fit using a bending power law model (see Section \ref{sec:xraytiming}).

\begin{figure}
\centering
\includegraphics[width=0.5\textwidth, keepaspectratio]{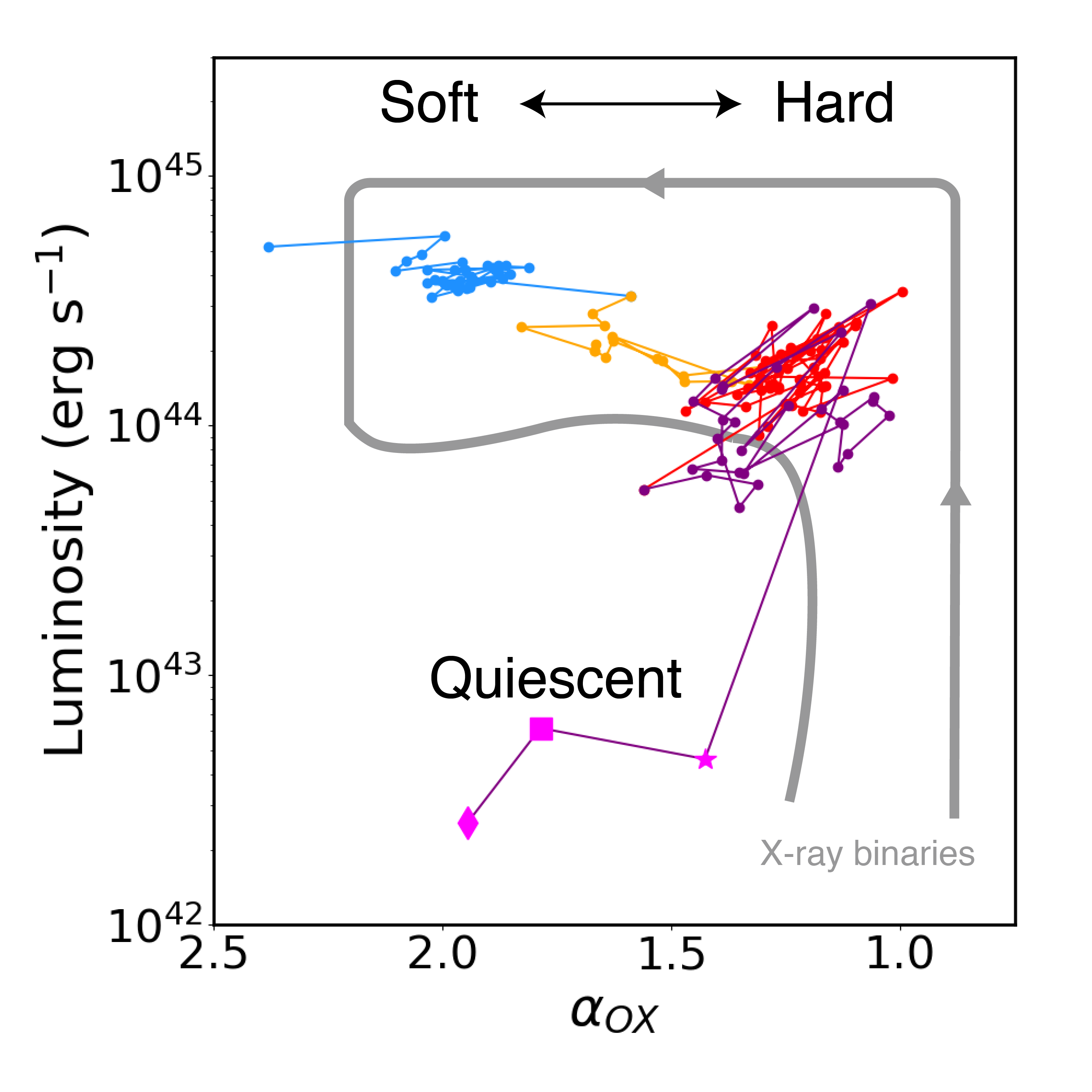}
\caption{AT2018fyk in the hardness - intensity diagram; hardness is represented by \alphaox. Note that we have reversed the horizontal axis to reproduce the classical HID. The grey bands show the typical turtle-head evolution of stellar-mass black holes in outburst. The TDE was discovered after peak light in the soft state; the observations missed the equivalent of the rising phase and hard to soft transition in X-ray binaries (segments marked with grey arrows).}
\label{fig:hid}
\end{figure}

Using these diagnostic tools, we investigate the properties of AT2018fyk in each of the phases defined in Figure \ref{fig:alphaox}, and find remarkable similarities to the properties of accreting stellar-mass black holes. The \alphaox evolution with \boleddfrac is strikingly similar to predictions from scaling a stellar-mass black hole outburst (\citealt{Sobolewska11a} and Fig. 3 of \citealt{Ruan19a}); we show a direct comparison with XRB outburst evolution across the hardness - intensity diagram (HID) in Figure \ref{fig:hid}, which further illustrates the similarities between these systems.

\subsection{Soft-to-hard accretion state transition}
At the time of discovery, we find AT2018fyk at high \alphaox, with an X-ray spectrum dominated by thermal emission with a temperature of $kT$ = 120--140 eV. The bolometric luminosity peaks around L$_{\rm bol}\approx$ 0.1L$_{\rm Edd}$ and is already declining, suggesting that we missed the rise to peak. The corona is present but weak, contributing only 5--25 \% of the X-ray flux and a few per cent of the bolometric luminosity. This is consistent with the power-law fraction observed in outbursting stellar-mass black holes in the soft state ($\sim 1-60 \%$; e.g. \citealt{Dunn10}), but inconsistent with the properties of soft state AGN where the corona is typically dominant (e.g. even for highly accreting narrow-line Seyfert 1 AGN the PL fraction is on average $70 \%$, see e.g \citealt{Vaughan99, Gliozzi20}; see also Section \ref{sec:agncomparison} for a more elaborate comparison to AGNs). The source is not detected at radio wavelengths (radio luminosity/X-ray luminosity = L$_{\rm radio}$ / L$_X \lesssim 10^{-6}$, \citealt{Wevers19b}), and we find no significant photometric variability on short timescales (frequencies $\gtrsim$10$^{-6}$ Hz; see Figure \ref{fig:psresp_comparison}). These properties are analogous with the soft states of stellar-mass black holes \citep{Homan05,Remillard06}.\\

When \boleddfrac drops below $\sim$ 0.05, roughly consistent with the typical transition luminosities of stellar mass black holes (ranging from a few to $\sim$10 \%, \citealt{Maccarone03}), AT2018fyk transitions into an intermediate state (phase B). Both the SED (i.e., \alphaox, top right panel of Figure \ref{fig:alphaox}) and the X-ray spectrum harden significantly over time (Figure \ref{fig:corfrac}). The observations show an increase in the X-ray power-law fraction (hardening), as well as an increasingly significant contribution of the power-law to the total bolometric luminosity. This is consistent with a scenario where the inner accretion disk becomes very cool or evaporates and the relative strength of the corona increases. 

When the source reaches phase C, \alphaox stabilises, and the power-law component dominates the X-ray spectrum (providing $>$50\% of the X-ray flux), and the X-ray flux varies dramatically. Rapid, large amplitude variability is present on both short (a few tens of minutes; bottom right panel of Figure \ref{fig:alphaox}) and long (days; top left panel of Figure \ref{fig:alphaox}) timescales. The onset of fast variability on roughly ten minute timescales is indicative of emission dominated by a compact emitting region, viz., an X-ray corona (see Section \ref{sec:xraytiming} and \ref{sec:timingdiscussion}). We note that the power-law contribution to the bolometric output in this state is similar to that seen in type 1 AGN with Eddington ratios $\sim$0.01 (\citealt{Lusso10}, see also Figure \ref{fig:plfrac_bol}).

The ratio of radio to X-ray luminosity is constrained to $\lesssim$ 4.5$\times$10$^{-7}$ in AT2018fyk. This rules out the appearance of a powerful radio jet, which is generally present in XRB hard states. Except for the absence of a radio jet, discussed in more detail in Section \ref{sec:radiodiscussion}, the rest of the properties are consistent with typical hard states found in XRBs and AGNs \citep{Homan05, Remillard06, Koerding06}. While the spectral hardening can, in principle, be due to variable absorption columns, the contemporaneous change in \alphaox as well as the change in timing properties cannot be explained in this way. Instead, this indicates that the disk transitioned from a thermal dominated soft state to a radiatively inefficient, power-law/non-thermal dominated hard state. 

X-ray flaring follows the soft-to-hard transition, when the thermal X-ray component has decreased in flux by $\sim$50 \%, while the power-law brightens by more than an order of magnitude. Rapid flares occur only in the X-ray band, indicating that they are confined to the hot, inner part of the accretion flow. While the X-ray spectral and photometric behaviour remain unchanged in the hard states (state C+D), the UV light curve in phase C is roughly constant whereas in phase D there are significant brightness variations (Figure \ref{fig:alphaox}, top left panel). The final phase (state D) of the UV light curve is remarkable, showing 2 cycles of {\it modulations} on $\sim$50 day timescales. There is no significant correlation between the UV and X-ray variability, which suggests physically distinct regions for these two phenomena. 

\subsection{Changes in the timing properties: appearance of a band-limited high frequency component}
\label{sec:timingdiscussion}
The soft state power spectrum (Figure \ref{fig:psresp_comparison}) is very different to that seen in the hard state, as unlike the hard state power spectrum, it can be well fitted (goodness-of-fit = 0.7) with a single steep power-law at low frequencies, with no power-spectral break to flatter slope required. The parameter confidence contours were obtained by assuming a low frequency slope of $\alpha_{\rm LF} = -1$ to have the best chance of matching the low frequency variability seen in the light curve, but a steeper slope extending to low frequencies is still preferred.

\begin{figure*}
\centering
\includegraphics[width=\textwidth, keepaspectratio]{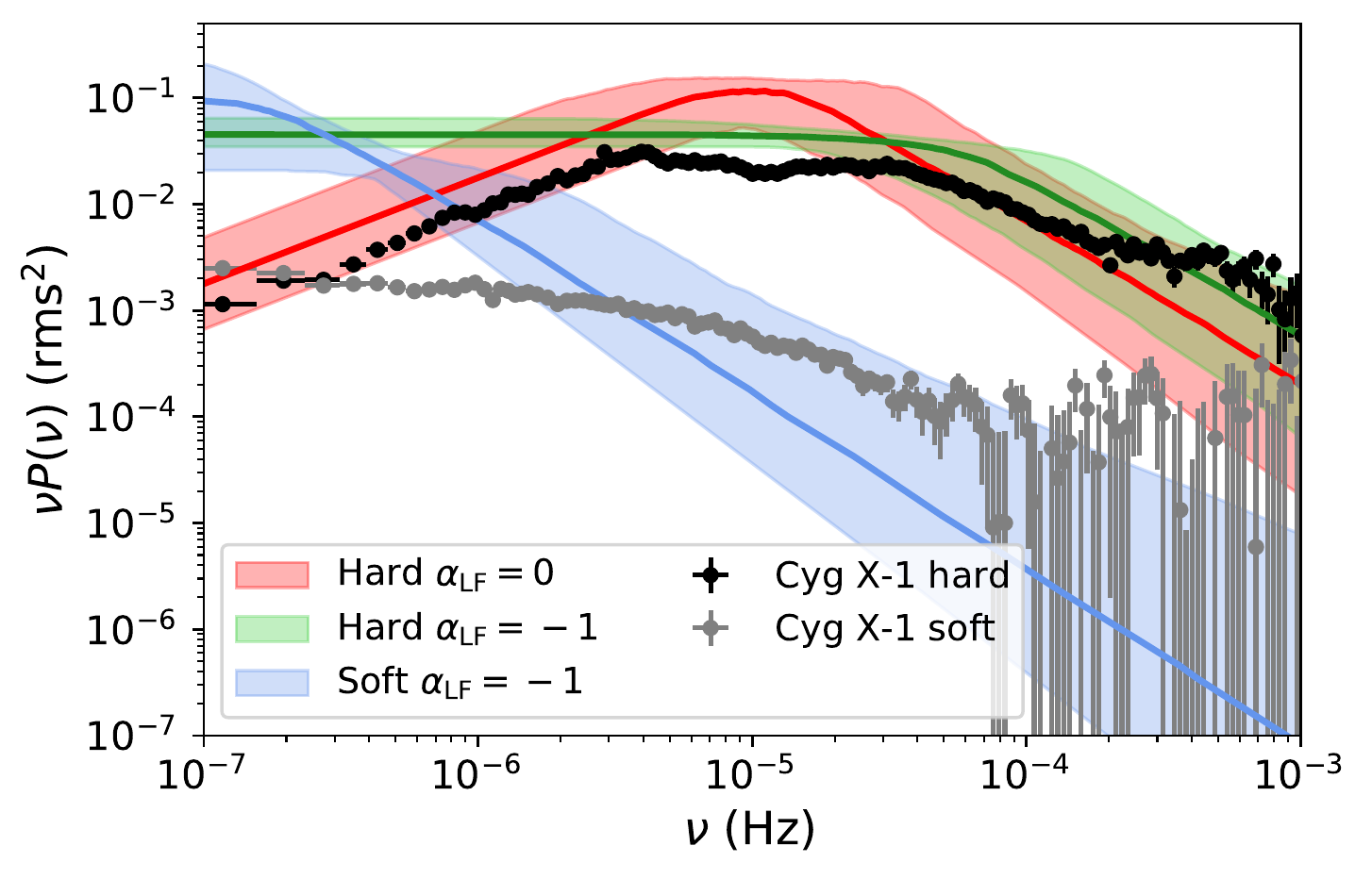}
\caption{Power density spectral models of the soft (state A) and hard (state C+D) states. The median (solid curve) and 95\% confidence intervals (shaded regions) of the models are shown. The confidence interval includes the effects of the uncertainty in high frequency slope and bend frequency. Sampling effects significantly distort the PDS from the {\it Swift} data, so we assume two different low frequency slopes (0 and --1) covering the range seen in AGN and X-ray binaries. Most of the variability/power comes from low frequencies in the soft state, while a high frequency component ($\gtrsim$10$^{-5}$ Hz) appears in the hard state and dominates the overall variability (see Table \ref{tab:psresp}). This supports the idea that a compact corona dominates emission in the hard state. The black and grey datapoints show the PDS of a well-observed X-ray binary, Cyg X-1, scaled up to a 10$^7$ M$_{\rm BH}$ black hole (see text).}
\label{fig:psresp_models}
\end{figure*}

The soft state shows a smaller fractional rms than the hard state, but this is concentrated entirely at low frequencies ($<10^{-6}$~Hz) while in the hard state a significant fraction of variability occurs above $10^{-5}$ Hz, as is apparent from the much more variable {\it XMM-Newton} light curve.

In Figure \ref{fig:psresp_comparison} we compare the soft and hard state power spectra directly. The lower bend frequency in the $\alpha_{LF} = 0$ case makes it a closer match to the low frequency constraint on the soft state bend and a more conservative comparison with the soft state power spectrum. However, the overall shapes of the distorted power spectra should not depend too strongly on the low frequency index, since both fit the data well. 

It is interesting to note that the low frequency power spectra for both soft and hard states are relatively consistent in amplitude, with a strong divergence in power at higher frequencies. Therefore, we can speculate that the main difference between the soft and hard state light curves is the addition of a band-limited high frequency component during the hard state. An additional broad high frequency component associated with the harder state is reminiscent of the transition between soft and hard states in X-ray binaries, where band-limited noise at higher frequencies is one of the first components associated with hard-intermediate states immediately prior to or after a transition to/from the soft state (e.g. see the power-spectral evolution presented in \citealt{Heil15}). 

We measure the break frequency of the red-noise PDS in the hard state and find log($\nu_{bend}$)$\approx$--4.5. This is on the lower end of the distribution of typical break frequencies found in AGN, which typically occur around $\nu_{bend}$ $\sim$ a few $\times$ 10$^{-4}$ Hz (although a lower frequency break similar to AT2018fyk has also been observed, \citealt{Gonzalez12}).
Assuming we can use the break frequency to black hole mass scaling relation derived for AGN \citep{Mchardy06} here, we derive a black hole mass of log(M$_{\rm BH}$) = 6.9--7.2 $\pm$ 0.55 M$_{\odot}$. This is in agreement to within the uncertainties with the estimate from the M--$\sigma$ relation \citep{Wevers20}.

It is evident from the XMM2 light curve (Figure \ref{fig:alphaox}, bottom right panel) that the brightness sometimes changes by $\sim$50\% on timescales as short as 1250 seconds. Using light travel time arguments, this corresponds to a very compact emission region of $<$5 R$_g$ for the observed X-ray variability, assuming M$_{\rm BH}$ = 10$^{7.7\pm0.4}$ M$_{\odot}$ (with an upper limit of R$<$13 R$_g$, allowing for the uncertainty in the M$_{\rm BH}$ measurement). These results support a scenario where the fastest X-ray variability originates in a very compact X-ray corona.

For comparison with X-ray binary PDS, we used NICER observations of Cyg X-1 in a hard (ObsID 2636010101) and soft state (ObsID 1100320122), obtaining PDS in respectively the 2--10 keV band (representing the hard power-law component variability) and 0.6--1 keV band (representing the disk blackbody component variability). Although Cyg X-1 shows uncharacteristically high variability amplitude in its soft state compared to other black hole X-ray binaries ($\sim10$\% rms vs. $<1$\% rms, e.g. \citealt{Heil15}), the fact that it accretes from a stellar wind (and hence has a relatively small accretion disk) may make it a better comparison with AGN, and with TDEs in particular.

PDS characteristic time-scales scale inversely with black hole mass and linearly with accretion rate. Assuming the $\sim 10^{7}$~M$_{\odot}$ black hole mass we infer from the measured break time-scale of the hard state TDE power spectrum, and a black hole mass of $\simeq$20~$M_{\odot}$ for Cyg X-1 \citep{Miller-Jones21}, combined with the larger (factor 5) accretion rate of the TDE hard state compared to Cyg X-1, we scale the Cyg~X-1 PDS by a factor $10^{-5}$ in frequency for a better comparison with the soft and hard TDE PDS in Figure \ref{fig:psresp_models}. This figure illustrates the similarity between the X-ray binary and TDE PDS in the soft and hard states.

\subsection{Constraints on the presence of a jet}
\label{sec:radiodiscussion}
TDEs are, in principle, excellent opportunities to study jet formation, as the newly formed accretion flow settles into a steady state. Follow up campaigns at radio wavelengths are indeed revealing a growing population of radio-bright TDEs \citep{Alexander20}. It is interesting to note that several (though not all) radio-bright TDEs launch their non-relativistic jets/outflows already at early times, while they are in the soft state \citep{vanvelzen2016, Stein20}. This is at odds with the standard paradigm in XRBs, where the jet is quenched during disk dominated accretion states \citep{Corbel02, Russell11} and the transition of the accretion flow from the soft into the hard state is generally accompanied by the emergence of a compact radio jet \citep{Fender03}.

While our early observations of AT2018fyk are consistent with quenched jets, if the analogy holds we would expect the emergence of a radio counterpart following the soft-to-hard state transition. 
However, our radio observations, covering the soft, hard and quiescent states, indicate that no bright radio jet is launched. Using the mean X-ray luminosity (3$\times$10$^{43}$ erg s$^{-1}$) in the hard state, the fundamental plane \citep{Koerdingplane} predicts a radio luminosity of $\sim$few $\times$ 10$^{39}$ erg s$^{-1}$. Our radio non detections constrain the luminosity to at least two orders of magnitude lower. We do note that a black hole mass that is lower by about 1 dex ($\sim$10$^{6.7}$ M$_{\odot}$) would bring the radio upper limits in line with the scatter observed in the fundamental plane. 

While the presence of a correlation among radio luminosity, X-ray luminosity and black hole mass is well established for black holes accreting in a radiatively inefficient state \citep{Merloni03, Falcke04}, the situation for systems in the process of changing accretion state is less clear. 
We briefly speculate which factors might be important to explain the apparent discrepancy with the XRB paradigm. 

The emergence of a compact radio jet following the soft-to-hard state transition has been well documented in X-ray binaries \citep{Fender01}, although there is a delay of one to several weeks between the transition in the X-ray spectrum and the onset of a radio source \citep{Kalemci13, Vahdat18}. The cause of this delay is currently unclear, as is any potential scaling with black hole mass, because multiple conditions need to be satisfied to successfully launch a jet. \citep{Kalemci13} found that for compact jets to be produced in XRBs, the disk luminosity Eddington fraction needs to be below 10$^{-4}$. This condition is not fulfilled for AT2018fyk, where the bolometric disk emission Eddington ratio is $> 10^{-3}$; other factors may also play a role. Once a strong corona forms, magnetic fields need to be generated and transported, which is one plausible explanation for the delay seen in XRBs. Alternatively, the corona has to be large enough to efficiently collimate the outflow into a compact radio jet.

If launching a jet requires a significant magnetic field, this will take time to grow via the magnetorotational instability. In contrast to AT2018fyk, several TDEs in the soft state do launch powerful outflows \citep{vanvelzen2016, Alexander20}, suggesting that the jet launching mechanism may be driven by other factors \citep{Pasham18}, such as the availability of magnetic flux. Since the disrupted star is not expected to carry significant magnetic flux, the build-up through dynamos or magnetorotational instabilities may be too slow for most TDEs, unless a fossil magnetic flux reservoir is present \citep{Kelley14}. 
We note that the host galaxy spectra of several radio-bright TDEs show possible evidence for nuclear activity. The narrow emission lines, when plotted in a diagnostic emission line diagram, fall in the regions indicative of a low ionisation narrow emission line region (LINER), AGN or composite AGN/star formation \citep{Nikolajuk13, French17}. While LINER-like emission can also be excited by other emission mechanisms, if instead this indicates recent or on-going low level AGN activity, the pre-existing jet cone and/or relic magnetic fields may facilitate the renewed launching of an outflow. AT2018fyk, on the other hand, does not show indications of recent AGN or star forming activity, so the absence of relic magnetic fields may also help explain the lack of a radio jet.

\subsection{Hard state to quiescence transition}
The strong UV and X-ray variability ends with a transition into a different regime (the quiescent state, E), evidenced by a factor $\sim$15 drop in the UV luminosity and a factor $>$5000 in X-rays. As a result, the SED significantly softens, i.e., \alphaox increases, and the UV emission becomes relatively more important when compared to the hard state. 

These dramatic changes constitute the {\it second} accretion state transition. Analogous, although much less dramatic, behaviour is also seen in stellar-mass black holes, which move towards a baseline, quiescent state at very low accretion rates. For stellar-mass black holes this state is defined empirically by L$_X <$ 10$^{-5}$ L$_{\rm Edd}$ \citep{Plotkin13}, when the anti-correlation between the photon index of the X-ray spectrum and L$_X$ plateaus or inverts \citep{Homan13, Yang15}. Qualitatively similar behaviour is also observed in (changing-look) AGN \citep{Yang15, Ruan19a}. 

The transition towards quiescence in stellar-mass black holes and AGN is more gradual and much less dramatic than observed for AT2018fyk. This warrants a discussion of alternative scenarios (see Section \ref{sec:thermalinstability}). For now, we note an important difference between X-ray binaries and AGN, and TDEs: the fallback rate of material at the outer disk edge. In stellar-mass black holes and AGN there is a steady supply of material, whereas in TDEs the supply is expected to decrease rapidly with time (following a t$^{-5/3}$ power-law behaviour \citealt{Rees1988}). This may trigger much more dramatic changes in the structural properties of the accretion flow in TDEs.
We constrain L$_X < 2.6 \times 10^{-6}$ L$_{\rm Edd}$ for AT2018fyk, indicating that the accretion flow has indeed reached the quiescent state and the mass supply rate from the stellar debris has dropped to very low levels.

The increase in \alphaox (softening of emission) at low Eddington ratio has also been observed in stellar-mass black hole outbursts \citep{Homan13,Plotkin17},
as well as in a sample of changing-look AGN \citep{Ruan19a}. Such a softening cannot be explained by the progressive disk truncation/evaporation thought to be responsible for the hardening of \alphaox from the soft to the hard state \citep{Sobolewska11b}. Instead, it suggests that the X-ray component is dominated by an entirely different emission mechanism, although observations are sparse and this state is poorly understood \citep{Plotkin15}. 
Remarkably, the \alphaox value in the quiescent state (state E) is at least as high as that observed in the soft state (state A), indicating that the X-ray corona is extremely weak or possibly non existent. 

\subsection{Alternative explanation for the second state transition}
\label{sec:thermalinstability}
Several peculiar features in the hard state (specifically, state D), including those discussed above, warrant a closer look to potential alternatives for the standard accretion state transition scenario for the total collapse of the system.
Thermal and viscous instabilities are an inherent property of 1D accretion disk theory at sub-Eddington accretion rates \citep{Lightman74,Shakura76}.
In particular, \citet{Shen14} predict that a thermal instability occurs when the fallback rate drops below the advection dominated regime. When this happens, the disk rapidly transitions into a gas pressure dominated, radiatively cooled state. Furthermore, in the presence of a sufficiently high rate of material falling back onto the outer disk, they predict limit cycle oscillations between this gas pressure dominated and the initial radiation pressure dominated states. More specifically, changes to both the disk radius as well as thermal state in the presence of fallback are expected, where the addition of low angular momentum material leads to a shrinking of the disk radius. This increases the disk surface density, and can cause the disk to jump back to the advective state for brief periods of time. 

The observed behaviour in AT2018fyk shows some similarities to theoretical predictions by \citet{Shen14}. First, the predicted timescale for the instabilities to manifest is $\sim$ 1 year, as observed. Second, the blackbody radius inferred in Figure \ref{fig:bbrad} shows an overall decreasing trend over time, as expected. Some oscillations are observed after $\sim$300 days, marking the onset of the UV photometric variability. \citet{Shen14} predict that the thermal cycles will end when the mass fallback rate drops below a critical threshold, which typically occurs after a few (2--3) cycles. AT2018fyk goes through 2 oscillatory cycles with a typical timescale of $\sim$50 days before the UV component (the accretion disk) eventually decreases by a large factor. This scenario provides a self-consistent explanation of the long term behaviour, the UV variability, the blackbody radius evolution, the subsequent state transition and the absence of a radio jet, providing tantalising evidence that thermal instabilities might occur in nature.

\subsection{The unusually rapid evolution of the accretion flow}
AT2018fyk is only the fourth UV/optical discovered TDE that is persistently X-ray bright for $\sim$500 days, the others being ASASSN--14li  \citep{Holoien201614li}, ASASSN--15oi \citep{Gezari17} and AT2019azh \citep{Hinkle20}. It shows evidence for a much more rapid disk formation and subsequent evolution when compared to typical TDEs, including the early emergence of Fe\,\textsc{ii} emission lines \citep{Wevers19b} and a pronounced second maximum early in the light curve \citep{Dong16, Leloudas16}. The sudden drop in UV luminosity also occurs relatively early when compared to other TDEs, with \citet{vanvelzen2019} finding only 1 TDE (SDSS TDE1) that may have experienced a similar drop in UV brightness 5--10 years after disruption. Similarly, the sharp decrease of the X-ray emission occurs much more rapidly after disruption than observed in other TDEs \citep{Jonker20}.

These seemingly peculiar properties suggest that the post disruption evolution proceeds more rapidly than in the bulk of the TDE population. Such a rapid evolution is expected if the debris streams can violently self-collide upon their return to pericenter, efficiently removing orbital angular momentum and energy. It is, however, unclear (both theoretically and observationally) whether this circularisation phase can indeed happen efficiently in the majority of TDEs \citep{Ulmer1999, Shiokawa2015, vanvelzen2019}.

The dominant mechanism to achieve strong shocks between streams is relativistic apsidal precession, which becomes increasingly important as the black hole mass increases, or alternatively when the star penetrates deeply into the SMBH potential well (high $\beta$ = R$_{p}$/R$_{t}$). 
Both AT2018fyk and SDSS TDE1 have comparatively high inferred black hole masses in excess of 10$^7$ M$_{\odot}$ \citep{Wevers17, Wevers20} and are UV faint at late times. At the inferred black hole mass of 1--5$\times$10$^7$ M$_{\odot}$, relativistic effects are important for all stellar encounters \cite{Stone19}, providing a natural explanation for the rapid formation and evolution of the accretion flow, which shines brightly in soft X-rays and at UV wavelengths.
The rapid subsequent evolution is a natural consequence of a high black hole mass. The Eddington limit in AT2018fyk is $\sim$10-50 times higher than in other well-observed TDEs, which typically have correspondingly lower black hole masses \citep{Wevers17, Wevers19a, Mockler19}. These results indicate that black hole mass could be an important factor setting the timescales of the post-disruption debris evolution.

An alternative explanation for rapid evolution could be a particularly low reservoir of debris to accrete from, i.e., a comparatively low mass star. As a result, the system will run out of material more quickly. However, stars with a mass $\lesssim$ 0.4 M$_{\odot}$ would be swallowed whole by a non-rotating SMBH in AT2018fyk (e.g. \citealt{Stone19}), while a dimensionless spin of $a > 0.7$ is required to produce an observable disruption for a 0.1 M$_{\odot}$ star (e.g. \citealt{Leloudas16}). Therefore, given the large mass of the black hole in AT2018fyk, it is unlikely that the disrupted stellar mass is significantly smaller when compared to typical TDEs \citep{Mockler19, vanvelzen2019, Wen20}. 

A conservative lower limit for the accreted stellar mass can be obtained by making an assumption for the conversion of accretion power to radiation. We calculate the bolometric energy output by piece wise integration of the entire light curve to find a total radiated energy of E$_{\rm rad}$ = 9.4$ \pm$ 0.2 $\times$ 10$^{51}$ ergs. Assuming a standard radiative efficiency of $\eta = 0.1$ and taking into account that only half of the stellar debris is available for accretion, the corresponding stellar mass is M$_{\star}\approx$0.1 M$_{\odot}$. We consider this calculation a lower limit to the true stellar mass. First, because of the assumption of radiatively efficient accretion throughout the evolution, which may not be valid when the source transitions into the hard state, where we may have $\eta \ll 0.1$. This could lead to an increase in the estimate of M$_{\star}$ of a factor 2--3, accounting for the observed duration of the soft and hard states respectively. Second, there are no constraints on the potentially large bolometric correction (up to a factor 10 or more, \citet{vanvelzen2016b}) due to dust absorption and re-emission on sub-parsec scales, which is the dominant contributor in the uncertainty for the radiated energy.

\subsection{Comparison of the corona properties with AGN}
\label{sec:agncomparison}
\begin{figure}
\includegraphics[width=\linewidth, keepaspectratio]{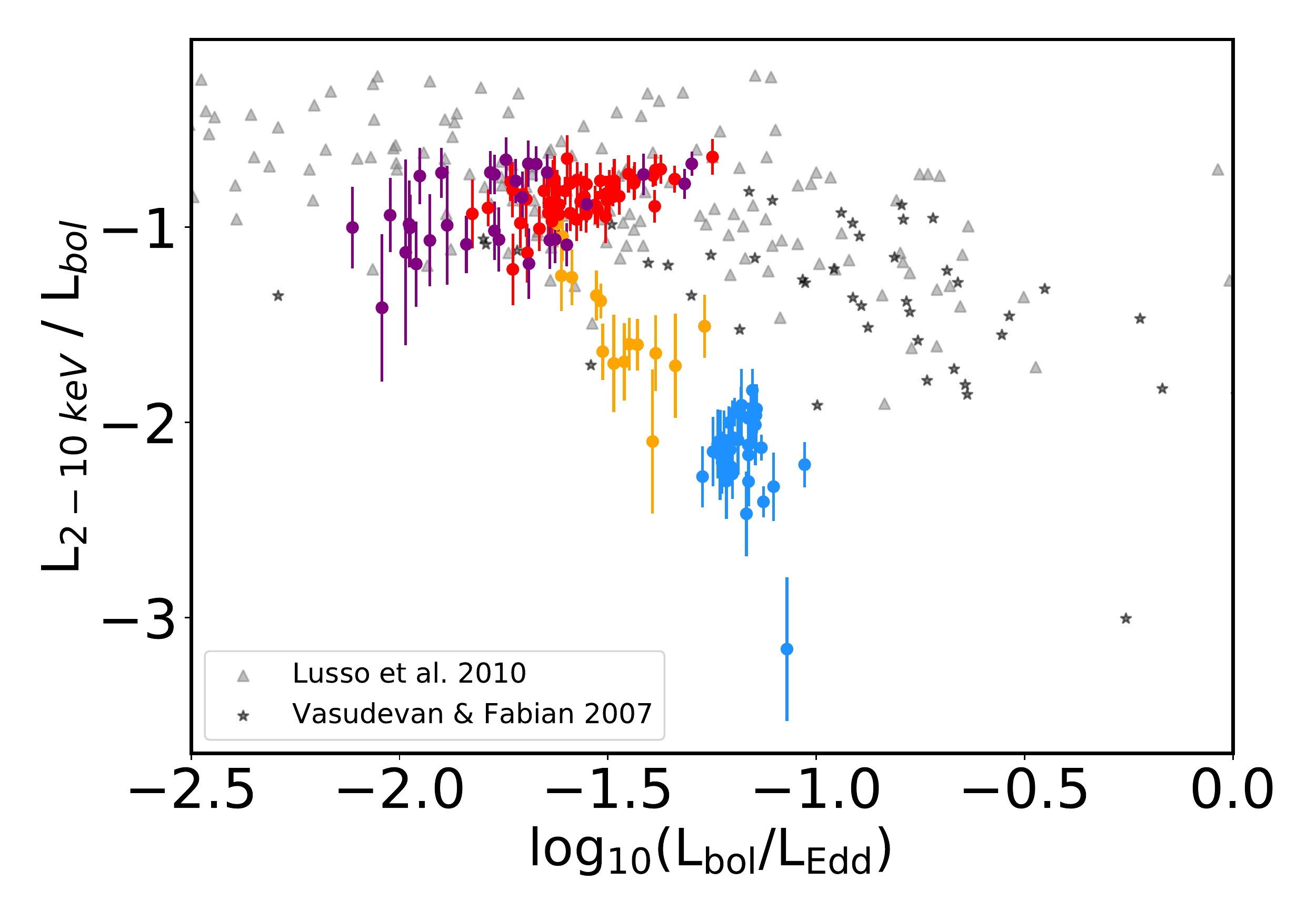}
\caption{Comparison of the evolution of the corona with AGN. Colour-coded filled circles show the evolution of the power law fraction to the bolometric emission as a function of the Eddington ratio. Grey and black triangles and stars represent AGN samples.}
\label{fig:plfrac_bol}
\end{figure}

It is still unclear whether the soft excess spectral component that is ubiquitously observed in AGN and TDEs is produced at the inner edge of the compact accretion disk, as an additional low temperature Comptonisation component, or elsewhere. This complicates the analogy and interpretation of the power-law fraction of emission as a proxy for accretion state, as is done for X-ray binaries.
To mitigate this uncertainty, an alternative approach that is directly analogous to the one employed for stellar-mass black hole systems is to consider not only the X-ray emission, but also include the UV emission where the accretion disk is expected to dominate for SMBHs. In Figure \ref{fig:corfrac} (right panel) we show the ratio of the 2--10 keV power-law emission to the total, bolometrically corrected (optical+UV+X-ray) emission (see Section \ref{sec:alphaox} for more details on how this correction is applied).
We illustrate how AT2018fyk compares to AGN samples in Figure \ref{fig:plfrac_bol}. The values in the hard state, at low Eddington ratios ($\sim$10$^{-2}$) cover a very similar parameter space to type 1 AGN (grey triangles, \citealt{Lusso10}) and other AGN samples including many NLS1 sources (black stars, \citealt{Vasudevan07}). However, in the soft state the power-law emission in AT2018fyk is on the low side compared to AGN with similar Eddington ratios. This could be resolved by taking into account the uncertainties on our black hole mass estimate $\sim$0.5 dex, and may indicate that the lower end of the mass range is favoured. Alternatively, we do not include the full energy ranges used to compute the bolometric corrections in the comparison samples of \citet{Lusso10} and \citet{Vasudevan07}; in particular, we don't have observational constraints for the 10--200 keV band. However, given the steep power-law indices we find, this should result in only small corrections, suggesting that the difference is physical in nature.

We recall that the power-law fraction in the soft state of AT2018fyk is comparable to the values seen in stellar-mass black holes, but much lower than in soft state AGN. The size of the accretion flow in TDEs (normalised to the black hole gravitational radius) is much more compact than in AGN. As a result, irradiation and/or radiation pressure are much higher in TDEs (a situation much more like X-ray binaries than in AGNs), which may prevent the formation of an efficient corona shortly after disruption, when the mass accretion rates are highest. Typical soft state AGN have been actively accreting material for long periods of time, and hence the corona has had time to build up and stabilise, whereas in the newly formed accretion flow following AT2018fyk we witnessed the formation of the corona in near real-time, strengthening as the source moves into the hard state. In the future, a larger TDE sample will help better understand the formation of the corona, and the nature (and role) of the poorly understood soft excess emission component seen in both TDEs and AGN.

\section{Summary}
\label{sec:summary}
We have presented a very rich, multi-wavelength set of observations of the tidal disruption event AT2018fyk, from peak light (at an Eddington ratio of $\sim$0.1) into quiescence (Eddington ratio $<$ 10$^{-3.4}$). 

We find that around peak light, the source exhibits several of the hallmark features of highly accreting supermassive black holes and stellar-mass black holes in the soft state, including a thermal-dominated X-ray spectrum, a soft (high \alphaox) SED, no bright radio emission and exclusively low-frequency ($<$10$^{-5}$ Hz) X-ray variability. After a transition phase, dramatic changes to these properties are observed. The power-law component becomes dominant in the X-ray spectrum, \alphaox hardens significantly and a band-limited high-frequency ($\sim$10$^{-3}$ Hz) time-variable component appears in the X-ray band. These changes are consistent with an accretion state transition from a soft to a hard state, and a strengthening of the X-ray corona. No bright radio jet accompanies this transition, marking an apparent deviation of the stellar-mass black hole accretion state paradigm. We discuss several plausible explanations for the absence of a radio jet around TDEs, and conclude that TDEs showing accretion state transitions provide a promising avenue to better understand the necessary conditions for successful jet formation around SMBHs with future observations. 

At late times, a second dramatic transition occurs, with the X-ray luminosity dropping by a factor $>$5000 and the UV luminosity decreasing by a factor $\sim$15. The SED significantly hardens during this transition, similar to the behaviour seen in changing-look AGN and other TDEs. This constitutes the second state transition, after which the source reaches the quiescent state at a low Eddington ratio $<$0.0004.

These transitions between the soft, hard and quiescent states, as well as the main drivers of disk (in)stabilities in accreting black holes remain poorly understood, especially so for SMBHs. This work shows that TDEs can provide a unique new opportunity to observationally constrain accretion state properties and transitions in individual SMBHs. In comparison to changing-look AGN \citep{Gezari17clagn, Ruan19a, Trakhtenbrot19, Frederick19} TDEs evolve faster, facilitating coordinated observational studies of the full evolution of accretion cycles and transitions. Furthermore, TDEs are readily discovered at increasing rates in wide-field optical and X-ray photometric surveys without the need for archival observations such as spectroscopy (to determine whether the source has changed state). We have demonstrated that a detailed characterisation of both the soft and hard states, as well as transition timescales and changes in the interplay between the different emitting regions, are possible. In summary, TDEs enable direct studies of the apparent scale invariance of accretion processes across seven orders of magnitude in black hole mass. In the future, statistical samples of TDEs with UV and radio monitoring observations, as well as soft and hard X-ray observations covering both short (stare) and long (monitoring) timescales will allow in-depth comparisons between stellar-mass and supermassive black hole accretion flow properties.

\acknowledgments
We thank D.J. Walton, W. Alston, A.C. Fabian, C. Reynolds, E. Coughlin, K. Hayasaki and M.I. Saladino for discussions, and the anonymous referee for constructive comments and suggestions.
We are grateful to the {\it Swift}, {\it XMM-Newton}, {\it NICER} and {\it Chandra} operation teams and PIs for awarding and scheduling the ToO/DDT observations. TW was funded in part by European Research Council grant 320360 and by European Commission grant 730980. SvV was supported by the James Arthur Postdoctoral Fellowship. This research was partially supported by the Australian Government through the Australian Research Council's Discovery Projects funding scheme (project DP200102471). Raw optical/UV/X-ray observations are available in the NASA/{Swift} archive (\url{http://heasarc.nasa.gov/docs/swift/archive}, Target Names: AT2018fyk, ASASSN-18UL); the XMM-Newton Science Archive (\url{http://nxsa.esac.esa.int}, obsIDs: 0831790201, 0853980201, 0854591401); and the Chandra data archive (\url{https://cxc.harvard.edu/cda/}, obsID: 23289). {\it NICER} data is publicly available through the HEASARC: \url{https://heasarc.gsfc.nasa.gov/cgi-bin/W3Browse/w3browse.pl}.

%

\vspace{5mm}
\facilities{XMM-Newton, Swift(XRT and UVOT), NICER, CXO, Magellan, ATCA}


\software{astropy \citep{astropy},  
          Heasoft \citep{heasoft}, 
          CIAO \citep{ciao}, 
          gPHOTON \citep{Million16}, 
          CASA \citep{McMullin2007}
          }


\clearpage

\bibliography{bibliography}{}

\begin{thebibliography}{}
\expandafter\ifx\csname natexlab\endcsname\relax\def\natexlab#1{#1}\fi
\providecommand{\url}[1]{\href{#1}{#1}}
\providecommand{\dodoi}[1]{doi:~\href{http://doi.org/#1}{\nolinkurl{#1}}}
\providecommand{\doeprint}[1]{\href{http://ascl.net/#1}{\nolinkurl{http://ascl.net/#1}}}
\providecommand{\doarXiv}[1]{\href{https://arxiv.org/abs/#1}{\nolinkurl{https://arxiv.org/abs/#1}}}

\bibitem[{{Abramowicz} \& {Fragile}(2013)}]{Abramowicz13}
{Abramowicz}, M.~A., \& {Fragile}, P.~C. 2013, Living Reviews in Relativity,
  16, 1, \dodoi{10.12942/lrr-2013-1}

\bibitem[{{Alexander} {et~al.}(2020){Alexander}, {van Velzen}, {Horesh}, \&
  {Zauderer}}]{Alexander20}
{Alexander}, K.~D., {van Velzen}, S., {Horesh}, A., \& {Zauderer}, B.~A. 2020,
  \ssr, 216, 81, \dodoi{10.1007/s11214-020-00702-w}

\bibitem[{{Arnaud}(1996)}]{Arnaud96}
{Arnaud}, K.~A. 1996, in Astronomical Society of the Pacific Conference Series,
  Vol. 101, Astronomical Data Analysis Software and Systems V, ed. G.~H.
  {Jacoby} \& J.~{Barnes}, 17

\bibitem[{{Astropy Collaboration} {et~al.}(2013){Astropy Collaboration},
  {Robitaille}, {Tollerud}, {Greenfield}, {Droettboom}, {Bray}, {Aldcroft},
  {Davis}, {Ginsburg}, {Price-Whelan}, {Kerzendorf}, {Conley}, {Crighton},
  {Barbary}, {Muna}, {Ferguson}, {Grollier}, {Parikh}, {Nair}, {Unther},
  {Deil}, {Woillez}, {Conseil}, {Kramer}, {Turner}, {Singer}, {Fox}, {Weaver},
  {Zabalza}, {Edwards}, {Azalee Bostroem}, {Burke}, {Casey}, {Crawford},
  {Dencheva}, {Ely}, {Jenness}, {Labrie}, {Lim}, {Pierfederici}, {Pontzen},
  {Ptak}, {Refsdal}, {Servillat}, \& {Streicher}}]{astropy}
{Astropy Collaboration}, {Robitaille}, T.~P., {Tollerud}, E.~J., {et~al.} 2013,
  \aap, 558, A33, \dodoi{10.1051/0004-6361/201322068}

\bibitem[{{Calzetti} {et~al.}(2000){Calzetti}, {Armus}, {Bohlin}, {Kinney},
  {Koornneef}, \& {Storchi-Bergmann}}]{Calzetti2000a}
{Calzetti}, D., {Armus}, L., {Bohlin}, R.~C., {et~al.} 2000, \apj, 533, 682,
  \dodoi{10.1086/308692}

\bibitem[{{Cash}(1979)}]{Cash79}
{Cash}, W. 1979, \apj, 228, 939, \dodoi{10.1086/156922}

\bibitem[{{Conroy} {et~al.}(2009){Conroy}, {Gunn}, \& {White}}]{Conroy09}
{Conroy}, C., {Gunn}, J.~E., \& {White}, M. 2009, \apj, 699, 486,
  \dodoi{10.1088/0004-637X/699/1/486}

\bibitem[{{Corbel} \& {Fender}(2002)}]{Corbel02}
{Corbel}, S., \& {Fender}, R.~P. 2002, \apjl, 573, L35, \dodoi{10.1086/341870}

\bibitem[{{Dey} {et~al.}(2019){Dey}, {Schlegel}, {Lang}, {Blum}, {Burleigh},
  {Fan}, {Findlay}, {Finkbeiner}, {Herrera}, {Juneau}, {Landriau}, {Levi},
  {McGreer}, {Meisner}, {Myers}, {Moustakas}, {Nugent}, {Patej}, {Schlafly},
  {Walker}, {Valdes}, {Weaver}, {Y{\`e}che}, {Zou}, {Zhou}, {Abareshi},
  {Abbott}, {Abolfathi}, {Aguilera}, {Alam}, {Allen}, {Alvarez}, {Annis},
  {Ansarinejad}, {Aubert}, {Beechert}, {Bell}, {BenZvi}, {Beutler}, {Bielby},
  {Bolton}, {Brice{\~n}o}, {Buckley-Geer}, {Butler}, {Calamida}, {Carlberg},
  {Carter}, {Casas}, {Castander}, {Choi}, {Comparat}, {Cukanovaite}, {Delubac},
  {DeVries}, {Dey}, {Dhungana}, {Dickinson}, {Ding}, {Donaldson}, {Duan},
  {Duckworth}, {Eftekharzadeh}, {Eisenstein}, {Etourneau}, {Fagrelius},
  {Farihi}, {Fitzpatrick}, {Font-Ribera}, {Fulmer}, {G{\"a}nsicke},
  {Gaztanaga}, {George}, {Gerdes}, {Gontcho}, {Gorgoni}, {Green}, {Guy},
  {Harmer}, {Hernandez}, {Honscheid}, {Huang}, {James}, {Jannuzi}, {Jiang},
  {Joyce}, {Karcher}, {Karkar}, {Kehoe}, {Kneib}, {Kueter-Young}, {Lan},
  {Lauer}, {Le Guillou}, {Le Van Suu}, {Lee}, {Lesser}, {Perreault Levasseur},
  {Li}, {Mann}, {Marshall}, {Mart{\'\i}nez-V{\'a}zquez}, {Martini}, {du Mas des
  Bourboux}, {McManus}, {Meier}, {M{\'e}nard}, {Metcalfe},
  {Mu{\~n}oz-Guti{\'e}rrez}, {Najita}, {Napier}, {Narayan}, {Newman}, {Nie},
  {Nord}, {Norman}, {Olsen}, {Paat}, {Palanque-Delabrouille}, {Peng},
  {Poppett}, {Poremba}, {Prakash}, {Rabinowitz}, {Raichoor}, {Rezaie},
  {Robertson}, {Roe}, {Ross}, {Ross}, {Rudnick}, {Safonova}, {Saha},
  {S{\'a}nchez}, {Savary}, {Schweiker}, {Scott}, {Seo}, {Shan}, {Silva},
  {Slepian}, {Soto}, {Sprayberry}, {Staten}, {Stillman}, {Stupak}, {Summers},
  {Sien Tie}, {Tirado}, {Vargas-Maga{\~n}a}, {Vivas}, {Wechsler}, {Williams},
  {Yang}, {Yang}, {Yapici}, {Zaritsky}, {Zenteno}, {Zhang}, {Zhang}, {Zhou}, \&
  {Zhou}}]{Dey19}
{Dey}, A., {Schlegel}, D.~J., {Lang}, D., {et~al.} 2019, \aj, 157, 168,
  \dodoi{10.3847/1538-3881/ab089d}

\bibitem[{{Done} \& {Gierli{\'n}ski}(2005)}]{Done05}
{Done}, C., \& {Gierli{\'n}ski}, M. 2005, \mnras, 364, 208,
  \dodoi{10.1111/j.1365-2966.2005.09555.x}

\bibitem[{{Dong} {et~al.}(2016){Dong}, {Shappee}, {Prieto}, {Jha}, {Stanek},
  {Holoien}, {Kochanek}, {Thompson}, {Morrell}, {Thompson}, {Basu}, {Beacom},
  {Bersier}, {Brimacombe}, {Brown}, {Bufano}, {Chen}, {Conseil}, {Danilet},
  {Falco}, {Grupe}, {Kiyota}, {Masi}, {Nicholls}, {Olivares E.}, {Pignata},
  {Pojmanski}, {Simonian}, {Szczygiel}, \& {Wo{\'z}niak}}]{Dong16}
{Dong}, S., {Shappee}, B.~J., {Prieto}, J.~L., {et~al.} 2016, Science, 351,
  257, \dodoi{10.1126/science.aac9613}

\bibitem[{{Dunn} {et~al.}(2010){Dunn}, {Fender}, {K{\"o}rding}, {Belloni}, \&
  {Cabanac}}]{Dunn10}
{Dunn}, R.~J.~H., {Fender}, R.~P., {K{\"o}rding}, E.~G., {Belloni}, T., \&
  {Cabanac}, C. 2010, \mnras, 403, 61, \dodoi{10.1111/j.1365-2966.2010.16114.x}

\bibitem[{{Falcke} {et~al.}(2004){Falcke}, {K{\"o}rding}, \&
  {Markoff}}]{Falcke04}
{Falcke}, H., {K{\"o}rding}, E., \& {Markoff}, S. 2004, \aap, 414, 895,
  \dodoi{10.1051/0004-6361:20031683}

\bibitem[{{Fender}(2001)}]{Fender01}
{Fender}, R.~P. 2001, \mnras, 322, 31, \dodoi{10.1046/j.1365-8711.2001.04080.x}

\bibitem[{{Fender} {et~al.}(2004){Fender}, {Belloni}, \& {Gallo}}]{Fender04}
{Fender}, R.~P., {Belloni}, T.~M., \& {Gallo}, E. 2004, \mnras, 355, 1105,
  \dodoi{10.1111/j.1365-2966.2004.08384.x}

\bibitem[{{Fender} {et~al.}(2003){Fender}, {Gallo}, \& {Jonker}}]{Fender03}
{Fender}, R.~P., {Gallo}, E., \& {Jonker}, P.~G. 2003, \mnras, 343, L99,
  \dodoi{10.1046/j.1365-8711.2003.06950.x}

\bibitem[{{Frederick} {et~al.}(2019){Frederick}, {Gezari}, {Graham}, {Cenko},
  {van Velzen}, {Stern}, {Blagorodnova}, {Kulkarni}, {Yan}, {De}, {Fremling},
  {Hung}, {Kara}, {Shupe}, {Ward}, {Bellm}, {Dekany}, {Duev}, {Feindt},
  {Giomi}, {Kupfer}, {Laher}, {Masci}, {Miller}, {Neill}, {Ngeow}, {Patterson},
  {Porter}, {Rusholme}, {Sollerman}, \& {Walters}}]{Frederick19}
{Frederick}, S., {Gezari}, S., {Graham}, M.~J., {et~al.} 2019, \apj, 883, 31,
  \dodoi{10.3847/1538-4357/ab3a38}

\bibitem[{{French} {et~al.}(2017){French}, {Arcavi}, \& {Zabludoff}}]{French17}
{French}, K.~D., {Arcavi}, I., \& {Zabludoff}, A. 2017, \apj, 835, 176,
  \dodoi{10.3847/1538-4357/835/2/176}

\bibitem[{{Fruscione} {et~al.}(2006){Fruscione}, {McDowell}, {Allen},
  {Brickhouse}, {Burke}, {Davis}, {Durham}, {Elvis}, {Galle}, {Harris},
  {Huenemoerder}, {Houck}, {Ishibashi}, {Karovska}, {Nicastro}, {Noble},
  {Nowak}, {Primini}, {Siemiginowska}, {Smith}, \& {Wise}}]{ciao}
{Fruscione}, A., {McDowell}, J.~C., {Allen}, G.~E., {et~al.} 2006, in Society
  of Photo-Optical Instrumentation Engineers (SPIE) Conference Series, Vol.
  6270, Society of Photo-Optical Instrumentation Engineers (SPIE) Conference
  Series, ed. D.~R. {Silva} \& R.~E. {Doxsey}, 62701V,
  \dodoi{10.1117/12.671760}

\bibitem[{{Gendreau} {et~al.}(2016){Gendreau}, {Arzoumanian}, {Adkins},
  {Albert}, {Anders}, {Aylward}, {Baker}, {Balsamo}, {Bamford}, {Benegalrao},
  {Berry}, {Bhalwani}, {Black}, {Blaurock}, {Bronke}, {Brown}, {Budinoff},
  {Cantwell}, {Cazeau}, {Chen}, {Clement}, {Colangelo}, {Coleman},
  {Coopersmith}, {Dehaven}, {Doty}, {Egan}, {Enoto}, {Fan}, {Ferro}, {Foster},
  {Galassi}, {Gallo}, {Green}, {Grosh}, {Ha}, {Hasouneh}, {Heefner}, {Hestnes},
  {Hoge}, {Jacobs}, {J{\o}rgensen}, {Kaiser}, {Kellogg}, {Kenyon}, {Koenecke},
  {Kozon}, {LaMarr}, {Lambertson}, {Larson}, {Lentine}, {Lewis}, {Lilly},
  {Liu}, {Malonis}, {Manthripragada}, {Markwardt}, {Matonak}, {Mcginnis},
  {Miller}, {Mitchell}, {Mitchell}, {Mohammed}, {Monroe}, {Montt de Garcia},
  {Mul{\'e}}, {Nagao}, {Ngo}, {Norris}, {Norwood}, {Novotka}, {Okajima},
  {Olsen}, {Onyeachu}, {Orosco}, {Peterson}, {Pevear}, {Pham}, {Pollard},
  {Pope}, {Powers}, {Powers}, {Price}, {Prigozhin}, {Ramirez}, {Reid},
  {Remillard}, {Rogstad}, {Rosecrans}, {Rowe}, {Sager}, {Sanders}, {Savadkin},
  {Saylor}, {Schaeffer}, {Schweiss}, {Semper}, {Serlemitsos}, {Shackelford},
  {Soong}, {Struebel}, {Vezie}, {Villasenor}, {Winternitz}, {Wofford},
  {Wright}, {Yang}, \& {Yu}}]{Gendreau16}
{Gendreau}, K.~C., {Arzoumanian}, Z., {Adkins}, P.~W., {et~al.} 2016, in
  Society of Photo-Optical Instrumentation Engineers (SPIE) Conference Series,
  Vol. 9905, \procspie, 99051H, \dodoi{10.1117/12.2231304}

\bibitem[{{Gezari} {et~al.}(2017{\natexlab{a}}){Gezari}, {Cenko}, \&
  {Arcavi}}]{Gezari17}
{Gezari}, S., {Cenko}, S.~B., \& {Arcavi}, I. 2017{\natexlab{a}}, \apj, 851,
  L47, \dodoi{10.3847/2041-8213/aaa0c2}

\bibitem[{{Gezari} {et~al.}(2017{\natexlab{b}}){Gezari}, {Hung}, {Cenko},
  {Blagorodnova}, {Yan}, {Kulkarni}, {Mooley}, {Kong}, {Cantwell}, {Yu}, {Cao},
  {Fremling}, {Neill}, {Ngeow}, {Nugent}, \& {Wozniak}}]{Gezari17clagn}
{Gezari}, S., {Hung}, T., {Cenko}, S.~B., {et~al.} 2017{\natexlab{b}}, \apj,
  835, 144, \dodoi{10.3847/1538-4357/835/2/144}

\bibitem[{{Giustini} {et~al.}(2020){Giustini}, {Miniutti}, \&
  {Saxton}}]{Giustini20}
{Giustini}, M., {Miniutti}, G., \& {Saxton}, R.~D. 2020, arXiv e-prints,
  arXiv:2002.08967.
\newblock \doarXiv{2002.08967}

\bibitem[{{Gliozzi} \& {Williams}(2020)}]{Gliozzi20}
{Gliozzi}, M., \& {Williams}, J.~K. 2020, \mnras, 491, 532,
  \dodoi{10.1093/mnras/stz3005}

\bibitem[{{Gonz{\'a}lez-Mart{\'\i}n} \& {Vaughan}(2012)}]{Gonzalez12}
{Gonz{\'a}lez-Mart{\'\i}n}, O., \& {Vaughan}, S. 2012, \aap, 544, A80,
  \dodoi{10.1051/0004-6361/201219008}

\bibitem[{{Heil} {et~al.}(2015){Heil}, {Uttley}, \& {Klein-Wolt}}]{Heil15}
{Heil}, L.~M., {Uttley}, P., \& {Klein-Wolt}, M. 2015, \mnras, 448, 3339,
  \dodoi{10.1093/mnras/stv191}

\bibitem[{{HI4PI Collaboration} {et~al.}(2016){HI4PI Collaboration}, {Ben
  Bekhti}, {Fl{\"o}er}, {Keller}, {Kerp}, {Lenz}, {Winkel}, {Bailin},
  {Calabretta}, {Dedes}, {Ford}, {Gibson}, {Haud}, {Janowiecki}, {Kalberla},
  {Lockman}, {McClure-Griffiths}, {Murphy}, {Nakanishi}, {Pisano}, \&
  {Staveley-Smith}}]{hi4pi}
{HI4PI Collaboration}, {Ben Bekhti}, N., {Fl{\"o}er}, L., {et~al.} 2016, \aap,
  594, A116, \dodoi{10.1051/0004-6361/201629178}

\bibitem[{{Hills}(1975)}]{Hills1975}
{Hills}, J.~G. 1975, \nat, 254, 295, \dodoi{10.1038/254295a0}

\bibitem[{{Hinkle} {et~al.}(2020){Hinkle}, {Holoien}, {Shappee}, {Auchettl},
  {Kochanek}, {Stanek}, {Payne}, \& {Thompson}}]{Hinkle20}
{Hinkle}, J.~T., {Holoien}, T. W.~S., {Shappee}, B.~J., {et~al.} 2020, arXiv
  e-prints, arXiv:2001.08215.
\newblock \doarXiv{2001.08215}

\bibitem[{{Holoien} {et~al.}(2016){Holoien}, {Kochanek}, {Prieto}, {Grupe},
  {Chen}, {Godoy-Rivera}, {Stanek}, {Shappee}, {Dong}, {Brown}, {Basu},
  {Beacom}, {Bersier}, {Brimacombe}, {Carlson}, {Falco}, {Johnston}, {Madore},
  {Pojmanski}, \& {Seibert}}]{Holoien201614li}
{Holoien}, T.~W.~S., {Kochanek}, C.~S., {Prieto}, J.~L., {et~al.} 2016, \mnras,
  463, 3813, \dodoi{10.1093/mnras/stw2272}

\bibitem[{{Homan} \& {Belloni}(2005)}]{Homan05}
{Homan}, J., \& {Belloni}, T. 2005, \apss, 300, 107,
  \dodoi{10.1007/s10509-005-1197-4}

\bibitem[{{Homan} {et~al.}(2013){Homan}, {Fridriksson}, {Jonker}, {Russell},
  {Gallo}, {Kuulkers}, {Rea}, \& {Altamirano}}]{Homan13}
{Homan}, J., {Fridriksson}, J.~K., {Jonker}, P.~G., {et~al.} 2013, \apj, 775,
  9, \dodoi{10.1088/0004-637X/775/1/9}

\bibitem[{{Hung} {et~al.}(2017){Hung}, {Gezari}, {Blagorodnova}, {Roth},
  {Cenko}, {Kulkarni}, {Horesh}, {Arcavi}, {McCully}, {Yan}, {Lunnan},
  {Fremling}, {Cao}, {Nugent}, \& {Wozniak}}]{Hung17}
{Hung}, T., {Gezari}, S., {Blagorodnova}, N., {et~al.} 2017, \apj, 842, 29,
  \dodoi{10.3847/1538-4357/aa7337}

\bibitem[{{Jansen} {et~al.}(2001){Jansen}, {Lumb}, {Altieri}, {Clavel}, {Ehle},
  {Erd}, {Gabriel}, {Guainazzi}, {Gondoin}, {Much}, {Munoz}, {Santos},
  {Schartel}, {Texier}, \& {Vacanti}}]{Jansen01}
{Jansen}, F., {Lumb}, D., {Altieri}, B., {et~al.} 2001, \aap, 365, L1,
  \dodoi{10.1051/0004-6361:20000036}

\bibitem[{{Jonker} {et~al.}(2020){Jonker}, {Stone}, {Generozov}, {Velzen}, \&
  {Metzger}}]{Jonker20}
{Jonker}, P.~G., {Stone}, N.~C., {Generozov}, A., {Velzen}, S.~v., \&
  {Metzger}, B. 2020, \apj, 889, 166, \dodoi{10.3847/1538-4357/ab659c}

\bibitem[{{Kalemci} {et~al.}(2013){Kalemci}, {Din{\c{c}}er}, {Tomsick},
  {Buxton}, {Bailyn}, \& {Chun}}]{Kalemci13}
{Kalemci}, E., {Din{\c{c}}er}, T., {Tomsick}, J.~A., {et~al.} 2013, \apj, 779,
  95, \dodoi{10.1088/0004-637X/779/2/95}

\bibitem[{{Kelley} {et~al.}(2014){Kelley}, {Tchekhovskoy}, \&
  {Narayan}}]{Kelley14}
{Kelley}, L.~Z., {Tchekhovskoy}, A., \& {Narayan}, R. 2014, \mnras, 445, 3919,
  \dodoi{10.1093/mnras/stu2041}

\bibitem[{{Komossa} {et~al.}(2004){Komossa}, {Halpern}, {Schartel}, {Hasinger},
  {Santos-Lleo}, \& {Predehl}}]{Komossa04}
{Komossa}, S., {Halpern}, J., {Schartel}, N., {et~al.} 2004, \apjl, 603, L17,
  \dodoi{10.1086/382046}

\bibitem[{{K{\"o}rding} {et~al.}(2006{\natexlab{a}}){K{\"o}rding}, {Falcke}, \&
  {Corbel}}]{Koerdingplane}
{K{\"o}rding}, E., {Falcke}, H., \& {Corbel}, S. 2006{\natexlab{a}}, \aap, 456,
  439, \dodoi{10.1051/0004-6361:20054144}

\bibitem[{{K{\"o}rding} {et~al.}(2006{\natexlab{b}}){K{\"o}rding}, {Jester}, \&
  {Fender}}]{Koerding06}
{K{\"o}rding}, E.~G., {Jester}, S., \& {Fender}, R. 2006{\natexlab{b}}, \mnras,
  372, 1366, \dodoi{10.1111/j.1365-2966.2006.10954.x}

\bibitem[{{Leloudas} {et~al.}(2016){Leloudas}, {Fraser}, {Stone}, {van Velzen},
  {Jonker}, {Arcavi}, {Fremling}, {Maund}, {Smartt}, {Kr{\`\i}hler},
  {Miller-Jones}, {Vreeswijk}, {Gal-Yam}, {Mazzali}, {De Cia}, {Howell},
  {Inserra}, {Patat}, {de Ugarte Postigo}, {Yaron}, {Ashall}, {Bar},
  {Campbell}, {Chen}, {Childress}, {Elias-Rosa}, {Harmanen}, {Hosseinzadeh},
  {Johansson}, {Kangas}, {Kankare}, {Kim}, {Kuncarayakti}, {Lyman}, {Magee},
  {Maguire}, {Malesani}, {Mattila}, {McCully}, {Nicholl}, {Prentice},
  {Romero-Ca{\~n}izales}, {Schulze}, {Smith}, {Sollerman}, {Sullivan},
  {Tucker}, {Valenti}, {Wheeler}, \& {Young}}]{Leloudas16}
{Leloudas}, G., {Fraser}, M., {Stone}, N.~C., {et~al.} 2016, Nature Astronomy,
  1, 0002, \dodoi{10.1038/s41550-016-0002}

\bibitem[{{Lightman} \& {Eardley}(1974)}]{Lightman74}
{Lightman}, A.~P., \& {Eardley}, D.~M. 1974, \apjl, 187, L1,
  \dodoi{10.1086/181377}

\bibitem[{{Lusso} {et~al.}(2010){Lusso}, {Comastri}, {Vignali}, {Zamorani},
  {Brusa}, {Gilli}, {Iwasawa}, {Salvato}, {Civano}, {Elvis}, {Merloni},
  {Bongiorno}, {Trump}, {Koekemoer}, {Schinnerer}, {Le Floc'h}, {Cappelluti},
  {Jahnke}, {Sargent}, {Silverman}, {Mainieri}, {Fiore}, {Bolzonella}, {Le
  F{\`e}vre}, {Garilli}, {Iovino}, {Kneib}, {Lamareille}, {Lilly}, {Mignoli},
  {Scodeggio}, \& {Vergani}}]{Lusso10}
{Lusso}, E., {Comastri}, A., {Vignali}, C., {et~al.} 2010, \aap, 512, A34,
  \dodoi{10.1051/0004-6361/200913298}

\bibitem[{{Maccarone} {et~al.}(2003){Maccarone}, {Gallo}, \&
  {Fender}}]{Maccarone03}
{Maccarone}, T.~J., {Gallo}, E., \& {Fender}, R. 2003, \mnras, 345, L19,
  \dodoi{10.1046/j.1365-8711.2003.07161.x}

\bibitem[{{Maksym} {et~al.}(2014){Maksym}, {Lin}, \& {Irwin}}]{Maksym14}
{Maksym}, W.~P., {Lin}, D., \& {Irwin}, J.~A. 2014, \apjl, 792, L29,
  \dodoi{10.1088/2041-8205/792/2/L29}

\bibitem[{{Marshall}(2015)}]{Marshall15}
{Marshall}, K. 2015, \apj, 810, 52, \dodoi{10.1088/0004-637X/810/1/52}

\bibitem[{{McElroy} {et~al.}(2016){McElroy}, {Husemann}, {Croom}, {Davis},
  {Bennert}, {Busch}, {Combes}, {Eckart}, {Perez-Torres}, {Powell},
  {Scharw{\"a}chter}, {Tremblay}, \& {Urrutia}}]{Mcelroy16}
{McElroy}, R.~E., {Husemann}, B., {Croom}, S.~M., {et~al.} 2016, \aap, 593, L8,
  \dodoi{10.1051/0004-6361/201629102}

\bibitem[{{McHardy} {et~al.}(2006){McHardy}, {Koerding}, {Knigge}, {Uttley}, \&
  {Fender}}]{Mchardy06}
{McHardy}, I.~M., {Koerding}, E., {Knigge}, C., {Uttley}, P., \& {Fender},
  R.~P. 2006, \nat, 444, 730, \dodoi{10.1038/nature05389}

\bibitem[{{McMullin} {et~al.}(2007){McMullin}, {Waters}, {Schiebel}, {Young},
  \& {Golap}}]{McMullin2007}
{McMullin}, J.~P., {Waters}, B., {Schiebel}, D., {Young}, W., \& {Golap}, K.
  2007, in Astronomical Society of the Pacific Conference Series, Vol. 376,
  Astronomical Data Analysis Software and Systems XVI, ed. R.~A. {Shaw},
  F.~{Hill}, \& D.~J. {Bell}, 127

\bibitem[{{Merloni} {et~al.}(2003){Merloni}, {Heinz}, \& {di
  Matteo}}]{Merloni03}
{Merloni}, A., {Heinz}, S., \& {di Matteo}, T. 2003, \mnras, 345, 1057,
  \dodoi{10.1046/j.1365-2966.2003.07017.x}

\bibitem[{{Miller-Jones} {et~al.}(2021){Miller-Jones}, {Bahramian}, {Orosz},
  {Mandel}, {Gou}, {Maccarone}, {Neijssel}, {Zhao}, {Zi{\'o}{\l}kowski},
  {Reid}, {Uttley}, {Zheng}, {Byun}, {Dodson}, {Grinberg}, {Jung}, {Kim},
  {Marcote}, {Markoff}, {Rioja}, {Rushton}, {Russell}, {Sivakoff}, {Tetarenko},
  {Tudose}, \& {Wilms}}]{Miller-Jones21}
{Miller-Jones}, J. C.~A., {Bahramian}, A., {Orosz}, J.~A., {et~al.} 2021, arXiv
  e-prints, arXiv:2102.09091.
\newblock \doarXiv{2102.09091}

\bibitem[{{Million} {et~al.}(2016){Million}, {Fleming}, {Shiao}, {Seibert},
  {Loyd}, {Tucker}, {Smith}, {Thompson}, \& {White}}]{Million16}
{Million}, C., {Fleming}, S.~W., {Shiao}, B., {et~al.} 2016, \apj, 833, 292,
  \dodoi{10.3847/1538-4357/833/2/292}

\bibitem[{{Miniutti} {et~al.}(2019){Miniutti}, {Saxton}, {Giustini}, {Alexand
  er}, {Fender}, {Heywood}, {Monageng}, {Coriat}, {Tzioumis}, {Read}, {Knigge},
  {Gandhi}, {Pretorius}, \& {Ag{\'\i}s-Gonz{\'a}lez}}]{Miniutti19}
{Miniutti}, G., {Saxton}, R.~D., {Giustini}, M., {et~al.} 2019, \nat, 573, 381,
  \dodoi{10.1038/s41586-019-1556-x}

\bibitem[{{Mockler} {et~al.}(2019){Mockler}, {Guillochon}, \&
  {Ramirez-Ruiz}}]{Mockler19}
{Mockler}, B., {Guillochon}, J., \& {Ramirez-Ruiz}, E. 2019, \apj, 872, 151,
  \dodoi{10.3847/1538-4357/ab010f}

\bibitem[{{Nasa High Energy Astrophysics Science Archive Research Center
  (Heasarc)}(2014)}]{heasoft}
{Nasa High Energy Astrophysics Science Archive Research Center (Heasarc)}.
  2014, {HEAsoft: Unified Release of FTOOLS and XANADU}.
\newblock \doeprint{1408.004}

\bibitem[{{Niko{\l}ajuk} \& {Walter}(2013)}]{Nikolajuk13}
{Niko{\l}ajuk}, M., \& {Walter}, R. 2013, \aap, 552, A75,
  \dodoi{10.1051/0004-6361/201220664}

\bibitem[{{Noda} \& {Done}(2018)}]{Noda18}
{Noda}, H., \& {Done}, C. 2018, \mnras, 480, 3898,
  \dodoi{10.1093/mnras/sty2032}

\bibitem[{{Parker} {et~al.}(2019){Parker}, {Schartel}, {Grupe}, {Komossa},
  {Harrison}, {Kollatschny}, {Mikula}, {Santos-Lle{\'o}}, \&
  {Tom{\'a}s}}]{Parker19}
{Parker}, M.~L., {Schartel}, N., {Grupe}, D., {et~al.} 2019, \mnras, 483, L88,
  \dodoi{10.1093/mnrasl/sly224}

\bibitem[{{Pasham} \& {van Velzen}(2018)}]{Pasham18}
{Pasham}, D.~R., \& {van Velzen}, S. 2018, \apj, 856, 1,
  \dodoi{10.3847/1538-4357/aab361}

\bibitem[{{Plotkin} {et~al.}(2013){Plotkin}, {Gallo}, \& {Jonker}}]{Plotkin13}
{Plotkin}, R.~M., {Gallo}, E., \& {Jonker}, P.~G. 2013, \apj, 773, 59,
  \dodoi{10.1088/0004-637X/773/1/59}

\bibitem[{{Plotkin} {et~al.}(2015){Plotkin}, {Gallo}, {Markoff}, {Homan},
  {Jonker}, {Miller-Jones}, {Russell}, \& {Drappeau}}]{Plotkin15}
{Plotkin}, R.~M., {Gallo}, E., {Markoff}, S., {et~al.} 2015, \mnras, 446, 4098,
  \dodoi{10.1093/mnras/stu2385}

\bibitem[{{Plotkin} {et~al.}(2017){Plotkin}, {Miller-Jones}, {Gallo}, {Jonker},
  {Homan}, {Tomsick}, {Kaaret}, {Russell}, {Heinz}, {Hodges-Kluck}, {Markoff},
  {Sivakoff}, {Altamirano}, \& {Neilsen}}]{Plotkin17}
{Plotkin}, R.~M., {Miller-Jones}, J.~C.~A., {Gallo}, E., {et~al.} 2017, \apj,
  834, 104, \dodoi{10.3847/1538-4357/834/2/104}

\bibitem[{{Rees}(1988)}]{Rees1988}
{Rees}, M.~J. 1988, \nat, 333, 523, \dodoi{10.1038/333523a0}

\bibitem[{{Remillard} \& {McClintock}(2006)}]{Remillard06}
{Remillard}, R.~A., \& {McClintock}, J.~E. 2006, \araa, 44, 49,
  \dodoi{10.1146/annurev.astro.44.051905.092532}

\bibitem[{{Roming} {et~al.}(2005){Roming}, {Kennedy}, {Mason}, {Nousek}, {Ahr},
  {Bingham}, {Broos}, {Carter}, {Hancock}, {Huckle}, {Hunsberger}, {Kawakami},
  {Killough}, {Koch}, {McLelland}, {Smith}, {Smith}, {Soto}, {Boyd},
  {Breeveld}, {Holland}, {Ivanushkina}, {Pryzby}, {Still}, \&
  {Stock}}]{Roming05}
{Roming}, P. W.~A., {Kennedy}, T.~E., {Mason}, K.~O., {et~al.} 2005, \ssr, 120,
  95, \dodoi{10.1007/s11214-005-5095-4}

\bibitem[{{Ruan} {et~al.}(2019){Ruan}, {Anderson}, {Eracleous}, {Green},
  {Haggard}, {MacLeod}, {Runnoe}, \& {Sobolewska}}]{Ruan19a}
{Ruan}, J.~J., {Anderson}, S.~F., {Eracleous}, M., {et~al.} 2019, \apj, 883,
  76, \dodoi{10.3847/1538-4357/ab3c1a}

\bibitem[{{Russell} {et~al.}(2011){Russell}, {Miller-Jones}, {Maccarone},
  {Yang}, {Fender}, \& {Lewis}}]{Russell11}
{Russell}, D.~M., {Miller-Jones}, J.~C.~A., {Maccarone}, T.~J., {et~al.} 2011,
  \apjl, 739, L19, \dodoi{10.1088/2041-8205/739/1/L19}

\bibitem[{{Shakura} \& {Sunyaev}(1976)}]{Shakura76}
{Shakura}, N.~I., \& {Sunyaev}, R.~A. 1976, \mnras, 175, 613,
  \dodoi{10.1093/mnras/175.3.613}

\bibitem[{{Shappee} {et~al.}(2014){Shappee}, {Prieto}, {Grupe}, {Kochanek},
  {Stanek}, {De Rosa}, {Mathur}, {Zu}, {Peterson}, {Pogge}, {Komossa}, {Im},
  {Jencson}, {Holoien}, {Basu}, {Beacom}, {Szczygie{\l}}, {Brimacombe},
  {Adams}, {Campillay}, {Choi}, {Contreras}, {Dietrich}, {Dubberley},
  {Elphick}, {Foale}, {Giustini}, {Gonzalez}, {Hawkins}, {Howell}, {Hsiao},
  {Koss}, {Leighly}, {Morrell}, {Mudd}, {Mullins}, {Nugent}, {Parrent},
  {Phillips}, {Pojmanski}, {Rosing}, {Ross}, {Sand}, {Terndrup}, {Valenti},
  {Walker}, \& {Yoon}}]{Shappee14}
{Shappee}, B.~J., {Prieto}, J.~L., {Grupe}, D., {et~al.} 2014, \apj, 788, 48,
  \dodoi{10.1088/0004-637X/788/1/48}

\bibitem[{{Shen} \& {Matzner}(2014)}]{Shen14}
{Shen}, R.-F., \& {Matzner}, C.~D. 2014, \apj, 784, 87,
  \dodoi{10.1088/0004-637X/784/2/87}

\bibitem[{{Shiokawa} {et~al.}(2015){Shiokawa}, {Krolik}, {Cheng}, {Piran}, \&
  {Noble}}]{Shiokawa2015}
{Shiokawa}, H., {Krolik}, J.~H., {Cheng}, R.~M., {Piran}, T., \& {Noble}, S.~C.
  2015, \apj, 804, 85, \dodoi{10.1088/0004-637X/804/2/85}

\bibitem[{{Sobolewska} {et~al.}(2011{\natexlab{a}}){Sobolewska}, {Papadakis},
  {Done}, \& {Malzac}}]{Sobolewska11b}
{Sobolewska}, M.~A., {Papadakis}, I.~E., {Done}, C., \& {Malzac}, J.
  2011{\natexlab{a}}, \mnras, 417, 280,
  \dodoi{10.1111/j.1365-2966.2011.19209.x}

\bibitem[{{Sobolewska} {et~al.}(2011{\natexlab{b}}){Sobolewska},
  {Siemiginowska}, \& {Gierli{\'n}ski}}]{Sobolewska11a}
{Sobolewska}, M.~A., {Siemiginowska}, A., \& {Gierli{\'n}ski}, M.
  2011{\natexlab{b}}, \mnras, 413, 2259,
  \dodoi{10.1111/j.1365-2966.2011.18302.x}

\bibitem[{{Stein} {et~al.}(2020){Stein}, {van Velzen}, {Kowalski},
  {Franckowiak}, {Gezari}, {Miller-Jones}, {Frederick}, {Sfaradi},
  {Bietenholz}, {Horesh}, {Fender}, {Garrappa}, {Ahumada}, {Andreoni},
  {Belicki}, {Bellm}, {B{\"o}ttcher}, {Brinnel}, {Burruss}, {Cenko},
  {Coughlin}, {Cunningham}, {Drake}, {Farrar}, {Feeney}, {Foley}, {Gal-Yam},
  {Golkhou}, {Goobar}, {Graham}, {Hammerstein}, {Helou}, {Hung}, {Kasliwal},
  {Kilpatrick}, {Kong}, {Kupfer}, {Laher}, {Mahabal}, {Masci}, {Necker},
  {Nordin}, {Perley}, {Rigault}, {Reusch}, {Rodriguez}, {Rojas-Bravo},
  {Rusholme}, {Shupe}, {Singer}, {Sollerman}, {Soumagnac}, {Stern}, {Taggart},
  {van Santen}, {Ward}, {Woudt}, \& {Yao}}]{Stein20}
{Stein}, R., {van Velzen}, S., {Kowalski}, M., {et~al.} 2020, arXiv e-prints,
  arXiv:2005.05340.
\newblock \doarXiv{2005.05340}

\bibitem[{{Stone} {et~al.}(2019){Stone}, {Kesden}, {Cheng}, \& {van
  Velzen}}]{Stone19}
{Stone}, N.~C., {Kesden}, M., {Cheng}, R.~M., \& {van Velzen}, S. 2019, General
  Relativity and Gravitation, 51, 30, \dodoi{10.1007/s10714-019-2510-9}

\bibitem[{{Tananbaum} {et~al.}(1979){Tananbaum}, {Avni}, {Branduardi}, {Elvis},
  {Fabbiano}, {Feigelson}, {Giacconi}, {Henry}, {Pye}, {Soltan}, \&
  {Zamorani}}]{Tananbaum79}
{Tananbaum}, H., {Avni}, Y., {Branduardi}, G., {et~al.} 1979, \apjl, 234, L9,
  \dodoi{10.1086/183100}

\bibitem[{{Trakhtenbrot} {et~al.}(2019){Trakhtenbrot}, {Arcavi}, {MacLeod},
  {Ricci}, {Kara}, {Graham}, {Stern}, {Harrison}, {Burke}, {Hiramatsu},
  {Hosseinzadeh}, {Howell}, {Smartt}, {Rest}, {Prieto}, {Shappee}, {Holoien},
  {Bersier}, {Filippenko}, {Brink}, {Zheng}, {Li}, {Remillard}, \&
  {Loewenstein}}]{Trakhtenbrot19}
{Trakhtenbrot}, B., {Arcavi}, I., {MacLeod}, C.~L., {et~al.} 2019, \apj, 883,
  94, \dodoi{10.3847/1538-4357/ab39e4}

\bibitem[{{Ulmer}(1999)}]{Ulmer1999}
{Ulmer}, A. 1999, \apj, 514, 180, \dodoi{10.1086/306909}

\bibitem[{{Uttley} {et~al.}(2002){Uttley}, {McHardy}, \&
  {Papadakis}}]{Uttley02}
{Uttley}, P., {McHardy}, I.~M., \& {Papadakis}, I.~E. 2002, \mnras, 332, 231,
  \dodoi{10.1046/j.1365-8711.2002.05298.x}

\bibitem[{{Vahdat Motlagh} {et~al.}(2019){Vahdat Motlagh}, {Kalemci}, \&
  {Maccarone}}]{Vahdat18}
{Vahdat Motlagh}, A., {Kalemci}, E., \& {Maccarone}, T.~J. 2019, \mnras, 485,
  2744, \dodoi{10.1093/mnras/stz569}

\bibitem[{{van Velzen} {et~al.}(2011){van Velzen}, {K{\"o}rding}, \&
  {Falcke}}]{vanvelzen11}
{van Velzen}, S., {K{\"o}rding}, E., \& {Falcke}, H. 2011, \mnras, 417, L51,
  \dodoi{10.1111/j.1745-3933.2011.01118.x}

\bibitem[{{van Velzen} {et~al.}(2016{\natexlab{a}}){van Velzen}, {Mendez},
  {Krolik}, \& {Gorjian}}]{vanvelzen2016b}
{van Velzen}, S., {Mendez}, A.~J., {Krolik}, J.~H., \& {Gorjian}, V.
  2016{\natexlab{a}}, \apj, 829, 19, \dodoi{10.3847/0004-637X/829/1/19}

\bibitem[{{van Velzen} {et~al.}(2019){van Velzen}, {Stone}, {Metzger},
  {Gezari}, {Brown}, \& {Fruchter}}]{vanvelzen2019}
{van Velzen}, S., {Stone}, N.~C., {Metzger}, B.~D., {et~al.} 2019, \apj, 878,
  82, \dodoi{10.3847/1538-4357/ab1844}

\bibitem[{{van Velzen} {et~al.}(2016{\natexlab{b}}){van Velzen}, {Anderson},
  {Stone}, {Fraser}, {Wevers}, {Metzger}, {Jonker}, {van der Horst}, {Staley},
  {Mendez}, {Miller-Jones}, {Hodgkin}, {Campbell}, \& {Fender}}]{vanvelzen2016}
{van Velzen}, S., {Anderson}, G.~E., {Stone}, N.~C., {et~al.}
  2016{\natexlab{b}}, Science, 351, 62, \dodoi{10.1126/science.aad1182}

\bibitem[{{van Velzen} {et~al.}(2020){van Velzen}, {Gezari}, {Hammerstein},
  {Roth}, {Frederick}, {Ward}, {Hung}, {Cenko}, {Stein}, {Perley}, {Taggart},
  {Sollerman}, {Andreoni}, {Bellm}, {Brinnel}, {De}, {Dekany}, {Feeney},
  {Foley}, {Fremling}, {Giomi}, {Golkhou}, {Ho}, {Kasliwal}, {Kilpatrick},
  {Kulkarni}, {Kupfer}, {Laher}, {Mahabal}, {Masci}, {Nordin}, {Riddle},
  {Rusholme}, {Sharma}, {van Santen}, {Shupe}, \& {Soumagnac}}]{vanvelzen20}
{van Velzen}, S., {Gezari}, S., {Hammerstein}, E., {et~al.} 2020, arXiv
  e-prints, arXiv:2001.01409.
\newblock \doarXiv{2001.01409}

\bibitem[{{Vasudevan} \& {Fabian}(2007)}]{Vasudevan07}
{Vasudevan}, R.~V., \& {Fabian}, A.~C. 2007, \mnras, 381, 1235,
  \dodoi{10.1111/j.1365-2966.2007.12328.x}

\bibitem[{{Vaughan} {et~al.}(1999){Vaughan}, {Reeves}, {Warwick}, \&
  {Edelson}}]{Vaughan99}
{Vaughan}, S., {Reeves}, J., {Warwick}, R., \& {Edelson}, R. 1999, \mnras, 309,
  113, \dodoi{10.1046/j.1365-8711.1999.02811.x}

\bibitem[{{Wen} {et~al.}(2020){Wen}, {Jonker}, {Stone}, {Zabludoff}, \&
  {Psaltis}}]{Wen20}
{Wen}, S., {Jonker}, P.~G., {Stone}, N.~C., {Zabludoff}, A.~I., \& {Psaltis},
  D. 2020, arXiv e-prints, arXiv:2003.12583.
\newblock \doarXiv{2003.12583}

\bibitem[{{Wevers}(2020)}]{Wevers20}
{Wevers}, T. 2020, \mnras, 497, L1, \dodoi{10.1093/mnrasl/slaa097}

\bibitem[{{Wevers} {et~al.}(2017){Wevers}, {van Velzen}, {Jonker}, {Stone},
  {Hung}, {Onori}, {Gezari}, \& {Blagorodnova}}]{Wevers17}
{Wevers}, T., {van Velzen}, S., {Jonker}, P.~G., {et~al.} 2017, \mnras, 471,
  1694, \dodoi{10.1093/mnras/stx1703}

\bibitem[{{Wevers} {et~al.}(2019{\natexlab{a}}){Wevers}, {Pasham}, {van
  Velzen}, {Leloudas}, {Schulze}, {Miller-Jones}, {Jonker}, {Gromadzki},
  {Kankare}, {Hodgkin}, {Wyrzykowski}, {Kostrzewa-Rutkowska}, {Moran},
  {Berton}, {Maguire}, {Onori}, {Mattila}, \& {Nicholl}}]{Wevers19b}
{Wevers}, T., {Pasham}, D.~R., {van Velzen}, S., {et~al.} 2019{\natexlab{a}},
  \mnras, 488, 4816, \dodoi{10.1093/mnras/stz1976}

\bibitem[{{Wevers} {et~al.}(2019{\natexlab{b}}){Wevers}, {Stone}, {van Velzen},
  {Jonker}, {Hung}, {Auchettl}, {Gezari}, {Onori}, {Mata S{\'a}nchez},
  {Kostrzewa-Rutkowska}, \& {Casares}}]{Wevers19a}
{Wevers}, T., {Stone}, N.~C., {van Velzen}, S., {et~al.} 2019{\natexlab{b}},
  \mnras, 487, 4136, \dodoi{10.1093/mnras/stz1602}

\bibitem[{{Wright} {et~al.}(2010){Wright}, {Eisenhardt}, {Mainzer}, {Ressler},
  {Cutri}, {Jarrett}, {Kirkpatrick}, {Padgett}, {McMillan}, {Skrutskie},
  {Stanford}, {Cohen}, {Walker}, {Mather}, {Leisawitz}, {Gautier}, {McLean},
  {Benford}, {Lonsdale}, {Blain}, {Mendez}, {Irace}, {Duval}, {Liu}, {Royer},
  {Heinrichsen}, {Howard}, {Shannon}, {Kendall}, {Walsh}, {Larsen}, {Cardon},
  {Schick}, {Schwalm}, {Abid}, {Fabinsky}, {Naes}, \& {Tsai}}]{Wright10}
{Wright}, E.~L., {Eisenhardt}, P. R.~M., {Mainzer}, A.~K., {et~al.} 2010, \aj,
  140, 1868, \dodoi{10.1088/0004-6256/140/6/1868}

\bibitem[{{Yang} {et~al.}(2015){Yang}, {Xie}, {Yuan}, {Zdziarski},
  {Gierli{\'n}ski}, {Ho}, \& {Yu}}]{Yang15}
{Yang}, Q.-X., {Xie}, F.-G., {Yuan}, F., {et~al.} 2015, \mnras, 447, 1692,
  \dodoi{10.1093/mnras/stu2571}

\end{thebibliography}
\bibliographystyle{aasjournal}



\end{document}